\input epsf
%
%
\def\bar#1{\overline{#1}}
\def\slash#1{\rlap{$#1$}/} 
\def\frac#1#2{{\textstyle{#1\over #2}}} 
\def\tr{\mathop{\rm tr}}
\def\Tr{\mathop{\rm Tr}}
\def\dsl{\,\raise.15ex\hbox{/}\mkern-13.5mu D} 
\def\delsl{\raise.15ex\hbox{/}\kern-.57em\partial}
\def\abs#1{\left| #1\right|}
\def\bra#1{\left\langle #1\right|}
\def\ket#1{\left| #1\right\rangle}
\def\CA{{\cal A}}\def\CD{{\cal D}}\def\CL{{\cal L}}\def\CO{{\cal O}}
\def\vev#1{\left\langle #1 \right\rangle}
\def\bfm#1{\rlap{$#1$}\mkern-0.75mu #1}
\def\lqcd{\Lambda_{\rm QCD}}
\def\leff{\CL_{\rm eff}}
\def\lchi{\Lambda_{\chi}}
\def\bfrac#1#2{ { #1 \over #2 }}
\def\boxtext#1{\vbox{\hrule\hbox{\vrule\kern5pt
       \vbox{\kern5pt{#1}\kern5pt}\kern5pt\vrule}\hrule}}
\def\tilde{\widetilde}

%
%
%
\catcode`\@=11 
%
\def\nolabels{\def\wrlabel##1{}\def\eqlabel##1{}\def\reflabel##1{}}
\def\writelabels{\def\wrlabel##1{\leavevmode\vadjust{\rlap{\smash%
{\line{{\escapechar=` \hfill\rlap{\sevenrm\hskip.03in\string##1}}}}}}}%
\def\eqlabel##1{{\escapechar-1\rlap{\sevenrm\hskip.05in\string##1}}}%
\def\reflabel##1{\noexpand\llap{\noexpand\sevenrm\string\string\string##1}}}
\nolabels
%
\global\newcount\secno \global\secno=0
\global\newcount\meqno \global\meqno=1
\def\newsec#1{\global\advance\secno by1\message{(\the\secno. #1)}
\global\subsecno=0\xdef\secsym{}
\bigbreak\bigskip\noindent{\bf\the\secno. #1}\writetoca{{\secsym} {#1}}
\par\nobreak\medskip\nobreak}
\xdef\secsym{}
\global\newcount\subsecno \global\subsecno=0
\def\subsec#1{\global\advance\subsecno by1\message{(\secsym\the\subsecno. #1)}
\bigbreak\noindent{\it\secsym\the\subsecno. #1}\writetoca{\string\quad
{\secsym\the\subsecno.} {#1}}\par\nobreak\medskip\nobreak}
\def\appendix#1#2{\global\meqno=1\global\subsecno=0\xdef\secsym{\hbox{#1.}}
\bigbreak\bigskip\noindent{\bf Appendix #1. #2}\message{(#1. #2)}
\writetoca{Appendix {#1.} {#2}}\par\nobreak\medskip\nobreak}
%
%
\def\eqnn#1{\xdef #1{(\secsym\the\meqno)}\writedef{#1\leftbracket#1}%
\global\advance\meqno by1\wrlabel#1}
\def\eqna#1{\xdef #1##1{\hbox{$(\secsym\the\meqno##1)$}}
\writedef{#1\numbersign1\leftbracket#1{\numbersign1}}%
\global\advance\meqno by1\wrlabel{#1$\{\}$}}
\def\eqn#1#2{\xdef #1{(\secsym\the\meqno)}\writedef{#1\leftbracket#1}%
\global\advance\meqno by1$$#2\eqno#1\eqlabel#1$$}
%
\newskip\footskip\footskip14pt plus 1pt minus 1pt 
\def\f@@t{\baselineskip\footskip\bgroup\aftergroup\@foot\let\next}
\setbox\strutbox=\hbox{\vrule height9.5pt depth4.5pt width0pt}
\global\newcount\ftno \global\ftno=0
\def\foot{\global\advance\ftno by1\footnote{$^{\the\ftno}$}}
%
\newwrite\ftfile
\def\footend{\def\foot{\global\advance\ftno by1\chardef\wfile=\ftfile
$^{\the\ftno}$\ifnum\ftno=1\immediate\openout\ftfile=foots.tmp\fi%
\immediate\write\ftfile{\noexpand\smallskip%
\noexpand\item{f\the\ftno:\ }\pctsign}\findarg}%
\def\footatend{\vfill\eject\immediate\closeout\ftfile{\parindent=20pt
\centerline{\bf Footnotes}\nobreak\bigskip\input foots.tmp }}}
\def\footatend{}
%
%
\global\newcount\refno \global\refno=1
\newwrite\rfile
\def\ref{{${}^{\the\refno}$}\nref}
\def\nref#1{\xdef#1{\the\refno}\writedef{#1\leftbracket#1}%
\ifnum\refno=1\immediate\openout\rfile=refs.tmp\fi
\global\advance\refno by1\chardef\wfile=\rfile\immediate
\write\rfile{\noexpand\item{#1.\ }\reflabel{#1\hskip.31in}\pctsign}\findarg}
\def\findarg#1#{\begingroup\obeylines\newlinechar=`\^^M\pass@rg}
{\obeylines\gdef\pass@rg#1{\writ@line\relax #1^^M\hbox{}^^M}%
\gdef\writ@line#1^^M{\expandafter\toks0\expandafter{\striprel@x #1}%
\edef\next{\the\toks0}\ifx\next\em@rk\let\next=\endgroup\else\ifx\next\empty%
\else\immediate\write\wfile{\the\toks0}\fi\let\next=\writ@line\fi\next\relax}}
\def\striprel@x#1{} \def\em@rk{\hbox{}}

\def\addref#1{\immediate\write\rfile{\noexpand\item{}#1}} 
\def\startrefs#1{\immediate\openout\rfile=refs.tmp\refno=#1}
\def\xref{\expandafter\xr@f}\def\xr@f[#1]{#1}
\def\refs#1{[\r@fs #1{\hbox{}}]}
\def\r@fs#1{\edef\next{#1}\ifx\next\em@rk\def\next{}\else
\ifx\next#1\xref #1\else#1\fi\let\next=\r@fs\fi\next}
%

%
\newwrite\ffile\global\newcount\figno \global\figno=1
\def\fig{\the\figno\nfig}
\def\nfig#1{\xdef#1{\the\figno}%
\writedef{#1\leftbracket \noexpand\the\figno}%
\ifnum\figno=1\immediate\openout\ffile=figs.tmp\fi\chardef\wfile=\ffile%
\immediate\write\ffile{\noexpand\medskip\noexpand\item{Fig.\ \the\figno. }
\reflabel{#1\hskip.55in}\pctsign}\global\advance\figno by1\findarg}
\def\xfig{\expandafter\xf@g}\def\xf@g \penalty\@M\ {}
\def\figs#1{\f@gs #1{\hbox{}}}
\def\f@gs#1{\edef\next{#1}\ifx\next\em@rk\def\next{}\else
\ifx\next#1\xfig #1\else#1\fi\let\next=\f@gs\fi\next}
\newwrite\lfile
{\escapechar-1\xdef\pctsign{\string\%}\xdef\leftbracket{\string\{}
\xdef\rightbracket{\string\}}\xdef\numbersign{\string\#}}

\def\writestop{\def\writestoppt{\immediate\write\lfile{\string\pageno%
\the\pageno\string\startrefs\leftbracket\the\refno\rightbracket%
\string\def\string\secsym\leftbracket\secsym\rightbracket%
\string\secno\the\secno\string\meqno\the\meqno}\immediate\closeout\lfile}}
\def\writestoppt{}\def\writedef#1{}
\def\seclab#1{\xdef #1{\the\secno}\writedef{#1\leftbracket#1}\wrlabel{#1=#1}}
\def\subseclab#1{\xdef #1{\secsym\the\subsecno}%
\writedef{#1\leftbracket#1}\wrlabel{#1=#1}}
\newwrite\tfile \def\writetoca#1{}
\def\leaderfill{\leaders\hbox to 1em{\hss.\hss}\hfill}
\def\writetoc{\immediate\openout\tfile=toc.tmp
   \def\writetoca##1{{\edef\next{\write\tfile{\noindent ##1
   \string\leaderfill {\noexpand\number\pageno} \par}}\next}}}

\catcode`@=12
%
%
%
%
\input temuphys.cmm
\contribution{Effective Field Theories}\footnote{}
{{\rm Lectures at the Schladming Winter School, March 1996, UCSD/PTH 96-04}}
\author{Aneesh V. Manohar}
\address{Physics Department, University of California, San Diego,
\hfill\break 9500 Gilman Drive, La Jolla, CA 92093, USA}
\abstract{These lectures introduce some of the basic ideas of effective field
theories. The topics discussed include: relevant and irrelevant operators and
scaling, renormalization in effective
field theories, decoupling of heavy particles, power counting, and
naive dimensional analysis.  Effective Lagrangians are used to study the
$\Delta S=2$ weak interactions and chiral perturbation theory.
}

\titlea{1}{Introduction}
An important idea that is implicit in all descriptions of physical
phenomena is that of an effective theory.  The basic premise of
effective theories is that dynamics at low energies (or large
distances) does not depend on the details of the dynamics at high
energies (or short distances).  As a result, low
energy physics can be described using an effective Lagrangian that contains
only a few
degrees of freedom, ignoring additional degrees of
freedom present at higher energies. One of the main purposes of these
lectures is to make these qualitative statements quantitative.

First a simple example: The energy levels of
the Hydrogen atom are calculated in textbooks using the Schr\"odinger
equation for an electron bound to a proton by a Coulomb potential.
To a good approximation,
the only properties of the proton that are relevant for the
computation are its mass and
charge.  An understanding of the quark substructure of the proton
(let alone quantum gravity) is not necessary to
compute the energy levels of the Hydrogen states.  This
is true provided an answer which has some
theoretical uncertainty is sufficient. A more accurate calculation
of the energy levels, for example including the hyperfine splitting,
requires that we also know that the proton has spin-1/2,
and a magnetic moment of 2.793 nuclear magnetons.  An even more
accurate calculation of the energy levels requires some knowledge of the
proton charge radius, etc.  More details of the proton structure
are needed as we require a more accurate answer for the energy levels.

When we discuss effective theories, we will frequently talk about
momentum scales characteristic of a given problem. The typical length
scale characteristic of the Hydrogen atom is the Bohr radius
$a_0=1/(m_{e}\alpha)$, and the typical momentum scale is of order
$\hbar/a_0\sim1/a_0 = m_{e}\alpha$,
using units in which $\hbar=1$. The typical energy scale
characteristic of Hydrogen is the Rydberg $\sim m_{e}\alpha^{2}$, and
the typical time scale is $1/(m_{e}\alpha^{2})$. The Hydrogen atom is
more complicated than many relativistic bound states because it has
two characteristic scales, $m_{e}\alpha$ and $m_{e}\alpha^{2}$. We can now
give a quantitative estimate of the error caused by neglected interactions: the
energy levels of Hydrogen can be computed by ignoring all dynamics on
momentum scales $\Lambda$ much larger than $m_{e}\alpha$,
with an error of order $m_{e} \alpha/\Lambda$. As the desired accuracy
increases, the scale $\Lambda$ of the interactions that can be ignored, also
increases.

The relevant interactions in an effective theory also depend on the
question being studied.  In the Hydrogen atom, the energy levels can
be computed to an accuracy $(m_{e}\alpha/M_{W})^{2}$ while ignoring
the weak interactions, but if we are interested in atomic parity
violation, the weak interactions are the leading contribution since
strong and electromagnetic interactions conserve parity.  Atomic
parity violation will still be a very small effect, because the weak
scale is much larger than the atomic scale.

An effective field theory describes low energy physics in terms of a few
parameters. These low energy parameters can be computed in terms of (hopefully
fewer) parameters in a more fundamental high energy theory. This computation
can
be done explicitly when the high energy theory is weakly coupled. In QED, for
example, one can predict low energy parameters such as the magnetic moment of
the electron which can be used in the Schr\"odinger equation. If the
high energy theory is strong coupled, as in QCD, one usually treats
the low energy parameters (such as the magnetic moment of the proton)
as free parameters that are fit to experiment. We will deal with both cases in
these lectures when we study the Fermi theory of weak interactions, and chiral
perturbation theory.

We have said that high energy dynamics can be ignored in
the study of processes at low energies. The precise form of this statement is
subtle. It is not true that parameters in
the high energy theory do not affect the low energy dynamics in any way.
The precise statement is that the
only effect of the high energy theory is to modify coupling constants in the
low energy theory, or to put symmetry constraints on the low energy theory.
The energy levels of Hydrogen should not depend on the masses of heavy
particles such as the top quark. This is not true: changing the top
quark mass while keeping the electromagnetic coupling constant at high
energies fixed, changes the electromagnetic coupling constant at low
energies,
\eqn\alphreln{
m_{t}{d\over dm_{t}} \left(1\over {\alpha}\right) = -{1\over 3 \pi}.
}
The proton mass also depends on the top quark mass,
\eqn\mproton{
m_{p} \propto m_{t}^{2/27}.
}
Despite this dependence, the value of $m_{t}$ is irrelevant for
studying the Hydrogen atom. The reason is that $\alpha$ and $m_{p}$ are
parameters of the Schr\"odinger equation for the Hydrogen atom. Fitting
to the observed energy levels determines the value of $\alpha$ at
low energies to be $1/137.036$, and the proton mass to be 938.27~MeV.
{\it The value of $m_{t}$ is irrelevant for atomic physics if the
Schr\"odinger equation is treated as a low energy theory whose
parameters $\alpha, m_{e}, m_{p}$ are determined from low energy
experiments}. The value of $m_{t}$ {\it is} relevant if one studies how
atomic physics changes as a function of $m_{t}$ while keeping the high energy
parameters constant.

High energy dynamics places non-trivial symmetry constraints on
a low energy effective theory. An interesting example of such a
constraint is the spin-statistics theorem. Non-relativistic quantum mechanics
is
a perfectly satisfactory theory, regardless of whether electrons are quantized
using Bose, Fermi or Boltzmann statistics.
However, a consistent relativistic
formulation of the theory requires that electrons obey Fermi
statistics, which is a constraint on non-relativistic quantum
mechanics that follows from causality in quantum electrodynamics.
The spin-statistics theorem is a statement about symmetry, and holds
regardless
of whether there is a simple connection between the high energy and low energy
theories. In low energy QCD, the spin-statistics theorem implies that baryons
are
fermions and mesons are bosons.

The effective field theory technique is powerful precisely because one can
compute low energy dynamics without any knowledge of the details of high
energy
interactions. This also has an unfortunate consequence --
information about high energy interactions cannot be obtained using low energy
measurements.
Luckily, the last statement is not quite true. There are some vestiges of the
high energy interactions in the symmetry constraints on the low energy theory,
and in small corrections to low energy dynamics. Thus high precision
low energy experiments can be used to probe high energy dynamics, and
provide an alternative to high energy experiments.

\titlea{2}{The Renormalization Group and Scaling}
Effective actions were used by Wilson, Fisher, and Kadanoff to study
critical phenomena in condensed matter systems, and many of the ideas
of effective theories were developed in this context. Consider the classic
 example of an Ising
spin system on a square
lattice with lattice spacing $a$, in an external magnetic field. The partition
function is
\eqn\IIi{
Z = \sum_{s_i=\pm} \exp\left[ { K \sum_{\vev{ij}} s_i s_j + B\sum_i s_i}
\right],}
where  $\vev{ij}$ is a sum over nearest neighbors. At a
second order phase transition, the correlation length $\xi$ of the system
becomes infinite. Intuitively, one expects that the properties of the
Ising system near its critical point should not depend on the details
of the system on the scale of the lattice spacing $a$. It took a
decade of inspired work to convert this intuitive statement into
equations.

To study the Ising model at its critical point,
it is not necessary to retain all the  information in the partition
function~\IIi. The idea  of Kadanoff was to reduce
the degrees of
freedom by introducing a block spin. Divide the lattice of spins into blocks
of four spins each (see Fig.~\fig\Kadanoff{The
Kadanoff block spin transformation. The four spins at the corner of each block
are replaced by an average spin at the center.}).
The block spin $s'$ is defined for
each block to be the average of the spins at the four corners of the
block
\eqn\block{
s_{B}' = {s_{B1}+s_{B2}+s_{B3}+s_{B4}\over4},
}
where $s_{B}'$ is the block spin for the block $B$, and $s_{Bi}$
are the original spins at the four corners of block $B$ .
One can write the partition function $Z$ as
\eqn\IIii{\eqalign{
Z &= \sum_{s_i=\pm} \exp \left[{ K \sum_{\vev{ij}} s_i s_j + B\sum_i s_i}
\right],\cr
&= \int \sum_{s_i=\pm}  \prod_B ds'_B
\delta\left(s'_B-{s_{B1}+s_{B2}+s_{B3}+s_{B4}\over4}\right)\cr
& \qquad\qquad\times
\exp \left[ { K \sum_{\vev{ij}} s_i s_j + B\sum_i s_i} \right],
}}
where the product is over all blocks $B$.
Performing the sum over $s_i$ leads to
\eqn\IIiii{
Z = \int \prod_B ds'_B\ e^{ S\left[s'_B\right] },
}
where
\eqn\IIiv{
e^{ S\left[s'_B\right] } = \sum_{s_i=\pm} \prod_B
\delta\left(s'_B-{s_{B1}+s_{B2}+s_{B3}+s_{B4}\over4}\right)
\exp \left[{ K \sum_{\vev{ij}} s_i s_j + B\sum_i s_i} \right].
}
This is an exact renormalization group transformation (called a Kadanoff
block spin transformation), that expresses the
partition function in terms of a new action $S\left[s'_B\right]$ with a
quarter the number of degrees of freedom and twice the lattice
spacing as the original action.

\begfig 0 cm
\centerline{\hbox{
\epsfxsize=6cm\epsffile{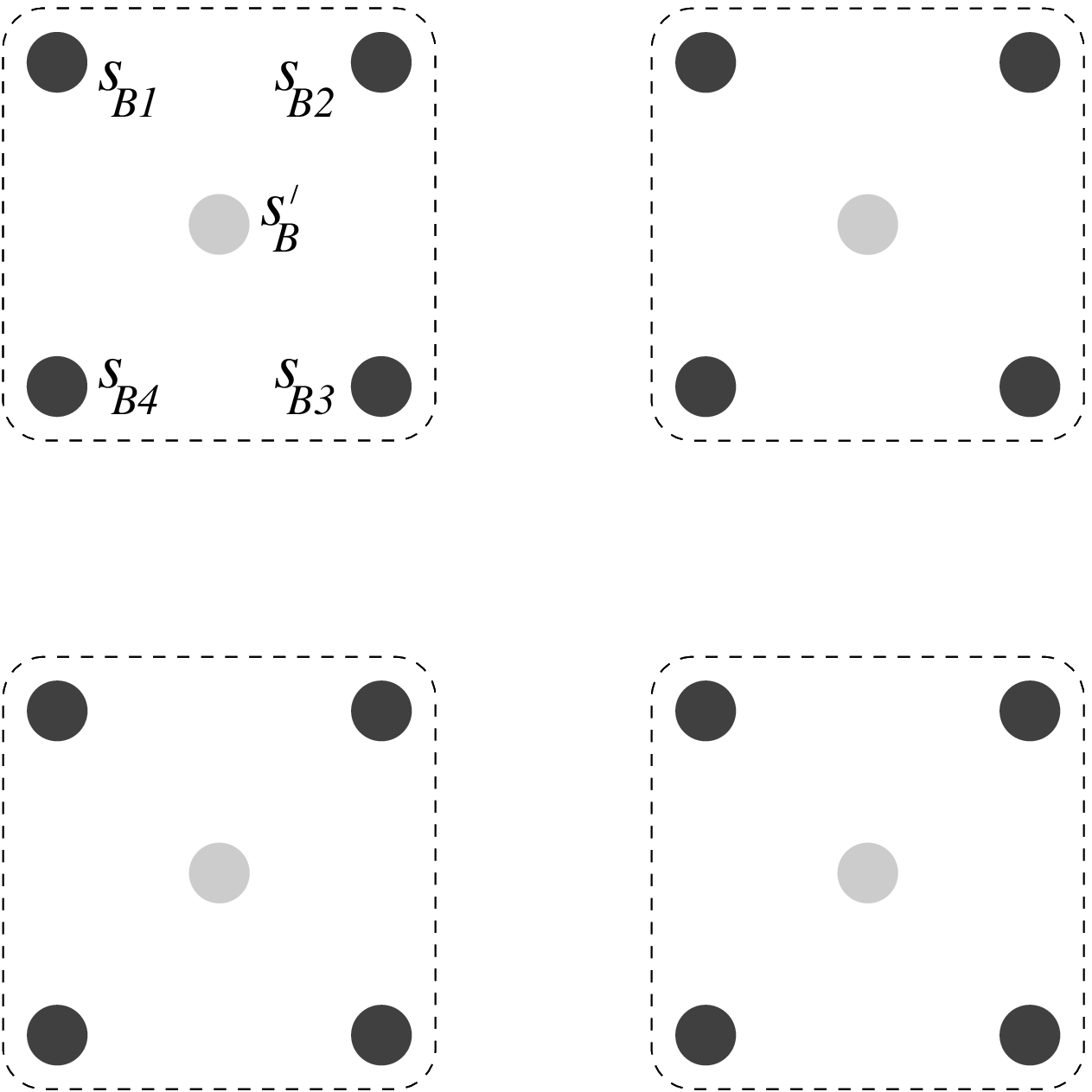}}}
\figure{\Kadanoff}{The
Kadanoff block spin transformation. The four spins at the corner of each block
are replaced by an average spin at the center}
\endfig

The new variable $s'_{B}$ is the average of four spins with values
$\pm 1$ \block, and can have the
values $\pm 1$, $\pm 1/2$, $0$. The new action $S\left[s'_B\right]$
is  much more complicated than the original action $K
\sum_{\vev{ij}} s_i s_j + B\sum_i s_i$, but can in principle be
computed using \IIiv.  Now repeat the block spin
transformation an infinite number of times.  At each step, the
number of degrees of freedom is reduced by four. The  block spin $s$
eventually becomes a continuous variable, which is  usually denoted by
$\phi$. The only problem is that the action becomes  more and more
complicated, and more and more non-local at each step.  This was the
difficulty that prevented the Kadanoff block spin method  from being
used for a long time. What is needed is a way to truncate  the
effective action in a systematic and controlled manner. It is also important
(particularly in
field theory) to have an effective action that is local.

A technical difficulty with the block spin transformation
is that it is discrete; it is much easier to deal with continuous
transformations. Wilson suggested studying the Ising
model in momentum space. The variables in momentum space are Fourier
transformed variables $s(k)$, where the momentum $k$ is restricted to
the Brillouin zone, $\abs{k} \le k_{\rm max}=\pi/a$. The Kadanoff
block spin transformation takes $a \rightarrow 2a
\rightarrow 4a$, etc., which corresponds to letting $ k_{\rm max}
\rightarrow k_{\rm max} /2 \rightarrow k_{\rm max} /4$, etc. $k$ is a
continuous variable, so instead consider decreasing $k_{\rm max}$
continuously. The
partition function transformation formula becomes (using $\phi, \Lambda$
instead of $s,k_{\rm max}$)
\eqn\intk{
Z = \int_{k \le \Lambda} \CD \phi_{k}\ e^{-S_{\Lambda}\left[ \phi_{k} \right]}
= \int_{k \le \Lambda'} \CD \phi'_{k}\ e^{-S'_{\Lambda'}\left[ \phi'_{k}
\right]},
}
which is the momentum space analog of \IIiv. The original action
$S_{\Lambda}\left[ \phi_{k}\right]$ contains all momentum modes up to
some maximum value $\Lambda$, whereas the new action $S'_{\Lambda'}\left[
\phi'_{k} \right]$ contains momentum modes up to $\Lambda'$, where
$\Lambda' < \Lambda$. The idea is to take $\Lambda' = \Lambda -
\delta \Lambda$ infinitesimally different from $\Lambda$, so that $S'$ is
infinitesimally different from $S$. In the limit $\delta \Lambda
\rightarrow 0$, the effective action satisfies a
differential equation,
\eqn\rgeqn{
{\partial S_{\Lambda}\over \partial \Lambda} = F\left[S_{\Lambda}\right],
}
where $F$ is a functional of the action that can be determined from
\intk. Think of $S_{\Lambda}$ as a set of actions, so that
\rgeqn\ gives the change in action as a function of cutoff. This
is usually referred to as the renormalization group flow of the action.
The action can be written as
\eqn\aexpand{
\sum_{i}c_{i}\ \CO_{i},
}
in terms of coefficients $c_{i}$
and some operator basis
$\CO_{i}$. The differential equation \rgeqn\ is then a differential
equation for the couplings,
\eqn\rgeqnii{
{\partial c_i\over \partial \Lambda} =
F\left[ \left\{ c_{i} \right\} \right],
}
so that the renormalization group equation gives a flow in coupling
constant space.

Finally, an extremely important point: the renormalization group
equations are obtained by integrating out variables with momenta
between $\Lambda - \delta \Lambda$ and $\Lambda$. There is both an
infrared ($\Lambda - \delta \Lambda$) and ultraviolet ($\Lambda$) cutoff on
the integration, so the
renormalization group equations are local and non-singular.

\titleb{}{Free Field Theory}
To explicitly study the renormalization group equations, it is
helpful to
consider first a free scalar field in $D$ dimensions, with action
\eqn\IVi{
S = \int d^D x \ {1\over2}\partial_\mu \phi \partial^\mu \phi - {1\over 2}
m^2 \phi^2.
}
The action $S$ is dimensionless, so the dimension of $\phi(x)$ is
determined from the kinetic term to be
$\left[\phi\right]=\left(D-2\right)/2$, and the dimension of $m^2$ is
$\left[m^2\right]=2$.

We would like to study correlation functions
\eqn\corri{
G_n\left(x_1,\ldots,x_n\right)=
\vev{ \phi\left(x_{1}\right) \ldots  \phi\left(x_{n}\right) }_{S},
}
computed using the action $S$ at long distances (i.e. low momentum).
It is convenient to make the
change of variables,
\eqn\IVii{
x = s x',\qquad \phi(x) = s^{(2-D)/2} \phi'(x'),
}
so that
\eqn\IViii{
S' = \int d^D x'\ {1\over 2}\partial'_\mu \phi'(x') \partial^{{}'\mu}\phi'(x')
-
{1\over 2} m^2 s^2 \phi'(x')^2.
}

Correlation functions of $\phi(x)$ with
action $S$ are related to correlation functions of $\phi'(x')$ with
action $S'$ by
\eqn\IViii{
\vev{\phi(s x_1)\ldots \phi(s x_n)}_S = s^{n(2-D)/2} \vev{\phi'(x_1)\ldots
\phi'(x_n)}_{S'}.
}
The long distance (low momentum) limit of correlation
functions with
action $S$ is
obtained by letting $s\rightarrow \infty$. These can be obtained by
studying correlation functions at a fixed distance (fixed momentum) of
the action $S'$ as $s \rightarrow \infty$. The mass term in $S'$ is
$s^{2}m^{2}$. Clearly, in the limit $s\rightarrow \infty$, the
mass term becomes more and more important. The mass $m^2$ is
called a relevant coupling, and dominates the long distance behavior
of the correlation functions. Equivalently, $\phi^{2}$ is called a
relevant operator.

What about integrating out
momentum shells to lower the cutoff? The original action had an
implicit cutoff $\Lambda$. We should have integrated out momentum
modes and lowered the cutoff to $\Lambda/s$, so that the rescaling
transformation \IVii\ restored the cutoff to its original value $\Lambda$.
In free field theory, there is no
coupling between the different modes. Thus integrating out a momentum
shell produces an overall multiplicative factor in $Z$, i.e. an
additive constant to the effective action. This shifts the
cosmological constant, but does not affect the dynamics of $\phi$.

\titleb{}{Interactions}
Next, add the interaction terms $\lambda \phi^4/4! + \lambda_6
\phi^6/6!$ to the free Lagrangian. The dimensions of the coefficients are
$\left[\lambda\right]=0$, $\left[\lambda_6\right]=-2$.  Rescaling the field as
before (and ignoring, for the moment, integrating out momenta between
$\Lambda$ and $\Lambda/s$) gives the
rescaled action
\eqn\IViv{
S' = \int d^D x'\ {1\over 2}\partial'_\mu \phi'\partial^{{}'\mu}\phi' -
{1\over 2} m^2 s^2 \phi'^2 - {\lambda\over 4!} \phi'^4 -
{\lambda_6\over 6! s^2} \phi'^6,
}
with the implicit rescaled cutoff $\Lambda$. In the limit
$s \rightarrow \infty$, the $\phi^6$ term vanishes as
$1/s^2$, so $\phi^6$ is called a irrelevant operator, and
$\lambda_{6}$ is called an irrelevant coupling. The $\phi^{4}$ term
remains unchanged under rescaling, so it is equally important at all
length scales. For this reason, $\phi^{4}$ is known as a marginal
operator, and $\lambda_{4}$ is called a marginal coupling.

In effective field theories, we are usually interested in studying the
dynamics at low energies, but not exactly at zero energy. For example, we will
be studying hadron dynamics at a scale of order 1~GeV, which is much smaller
than the weak interaction scale of $M_W \sim 80$~GeV. In this case, the scale
factor $s$ between the weak and strong scales is $s=80$, which is large but
finite. Irrelevant operators (despite their name) then produce small
corrections. In our scalar field theory example, the $\phi^6$ operator
produces corrections of order $1/s^2$, $\phi^8$ produces corrections of order
$1/s^4$, etc.

The alert reader will have noticed that the above results follow from
dimensional analysis. The counting can trivially be generalized to an
arbitrary  Lagrangian:
\medskip
\item{1.} Determine the canonical dimensions of the fields using the kinetic
term.
\item{2.} Determine the (mass) dimensions of all the couplings.
\item{3.} Terms with a coupling constant with dimension $d$ scale as
$s^d$, so that the coupling is relevant, irrelevant or marginal depending on
whether $d>0$, $d<0$ or $d=0$. Equivalently, the operator is relevant,
irrelevant, or marginal depending on whether its dimension is less
than, greater than, or equal to the space-time dimension $D$.
\item{4.}  To include all corrections up to order $1/s^r$, one should include
all operators with dimension $\le D+r$, i.e. all terms with coefficients of
dimension $\ge -r$.
\medskip
\noindent Let us now turn on the interactions. The first problem is that there
are divergences in the quantum theory. These are handled by the
standard regularization and renormalization procedure. In scalar
field theory, for example, one can introduce a cutoff $\Lambda$ to
regulate the functional integral. In the presence of a cutoff, the
relation \IViii\ between correlation functions becomes
\eqn\greln{
G_n \left(\left\{s x \right\}; m^{2}, \lambda_{4},\lambda_{6};
\Lambda\right) = s^{n\left(2-D\right)/2}
G_n\left(\left\{x \right\}; s^{2} m^{2}, \lambda_{4},s^{-2}\lambda_{6};
s \Lambda\right),
}
where we have explicitly included the cutoff dependence, and $\left\{x
\right\}$ denotes $x_1,\ldots,x_n$. The left
hand side is the desired correlation function. To get the infrared
behavior of the left hand side, we need to replace the cutoff
$s\Lambda$ by $\Lambda$ on the right hand side. This is the hard part
of the calculation which we have ignored so far,
but one that you have all seen before -- it is the standard renormalization
group equation of quantum field theory:
\eqn\rgfield{
\left[\Lambda {\partial \over \partial \Lambda} + \beta_{i}
{\partial \over \partial c_{i}} + n \gamma_\phi \right] G = 0.
}
Here $c_{i}$ are the couplings, $m^{2}, \lambda, \lambda_{6}$, etc., and the
$\beta$-functions $\beta_{i}$ and anomalous dimension $\gamma_\phi$ are
functions of $c_{i}$. The solution of this equation is also standard.
Define running couplings which are solutions of the differential
equation
\eqn\runc{
\Lambda {\partial \over \partial \Lambda} c_{i}\left(\Lambda
\right) = \beta_{i}\left(c_{i}\left(\Lambda \right)\right).
}
Then
\eqn\rgsoln{
G_n\left(\left\{ x \right\} ,c_{i}\left(\Lambda_{1}\right),\Lambda_{1}\right)=
e^{-n \int_{\Lambda_{1}}^{\Lambda_{2}}\gamma\left(\Lambda \right) d
\log \Lambda }\
G_n\left(\left\{ x \right\},c_{i}\left(\Lambda_{2}\right),\Lambda_{2}\right).
}
Equation~\greln\ can be combined with \rgsoln\ to give
\eqn\finalrg{
G\left(\left\{s x \right\},c_{i}\left(\Lambda \right),\Lambda \right)=
s^{n\left(2-D\right)/2}
e^{-n \int_{\Lambda}^{\Lambda/s}\gamma\left(\Lambda' \right) d
\log \Lambda' }
G\left(\left\{ x \right\},s^{d_{i}}c_{i}\left(\Lambda/s\right),\Lambda\right)
}
where $d_{i}$ is the dimension of coupling $c_{i}$. The only
difference from \IViii\ is the exponential prefactor, and that $c_{i}$ is now
the running coupling at $\Lambda/s$.

It is instructive to look at some examples of renormalization group equations
before continuing with our general analysis. In QCD, the renormalization group
equation for the dimensionless coupling constant $g$ is
\eqn\qcdrg{
\mu {\partial g\over \partial \mu} = - {g^3 \over 16 \pi^2} b_0 + \CO \left(
g^5 \right),
}
where $b_0=11 N_c/3 - 2 N_f/3$, $N_c$ is the number of colors, and $N_f$ is
the number of flavors. Equation~\qcdrg\ is the $\beta$-function in any mass
independent scheme, such as $\overline {MS}$, and $\mu$ is the dimensionful
parameter that plays the role of $\Lambda$ in such a scheme. The anomalous
dimension for a field (such as the fermion field $\psi$) has the form
\eqn\qcdad{
\gamma_\psi = \gamma_\psi^0 {g^2 \over 16 \pi^2} + \CO \left( g^4 \right).
}
Other operators added to the Lagrangian, such as four-Fermi weak decay
operators have renormalization group equations of the form
\eqn\qcdop{
\mu {\partial c_i\over \partial \mu} = \gamma_{ij}^0 {g^2 \over 16 \pi^2}c_j +
\CO \left( g^4 \right).
}
Let us neglect operator mixing for simplicity, so that $\gamma_{ij}$ is a
diagonal matrix, with elements $\gamma_i \delta_{ij}$. The solutions of the
renormalization group equations are\fonote{The general case where $\gamma_{ij}$
is not
diagonal can be solved by finding the eigenvalues and eigenvectors of
$\gamma_{ij}$.}
\eqn\rgsolni{\eqalign{
&{1\over \alpha_s \left( \mu_1 \right) }-
{1\over \alpha_s \left( \mu_2  \right) } = {b_0\over 2 \pi} \log { \mu_1 \over
\mu_2} ,\cr
&{ c_i \left( \mu_1 \right) \over c_i \left( \mu_2 \right) }
= \left[ { \alpha_s \left( \mu_1 \right) \over \alpha_s \left( \mu_1
\right)}\right]^{-\gamma_i^0 /2b_0} ,\cr
& \exp\left[-\int_{\mu_2}^{\mu_1} \gamma_\psi \left( \mu \right) d \log \mu
\right]
= \left[ { \alpha_s \left( \mu_1 \right) \over \alpha_s \left( \mu_2
\right)}\right]^{\gamma^0_\psi /2b_0}.
}}
These equations show that the quantum scaling behavior differs from the
classical one by logarithms.
A more interesting case is a field theory in which the $\beta$-function
for a dimensionless coupling such as $g$ has the form shown in
Fig.~\fig\zerobeta{An infrared stable fixed point of the $\beta$ function.}.
The renormalization group equation for $g$, \qcdrg, shows that $g
\rightarrow g^*$ as $\mu \rightarrow 0$ if $g$ starts out in some neighborhood
of $g^*$. For this reason $g^*$ is known as an attractive (or stable) infrared
fixed point for $g$. In this case, the renormalization group scaling in the
limit $s \rightarrow \infty$ is dominated by $g \approx g^*$, so that
\eqn\fixedrg{\eqalign{
\mu {\partial c_i\over \partial \mu} &= \gamma_{ij}\left( g^* \right) c_j, \cr
\gamma_\phi\left( \mu \right ) & \rightarrow \gamma_\phi\left( g^* \right ).
}}
Denote the fixed point values $\gamma_{ij}\left( g^* \right)$ and
$\gamma\left( g^* \right )$ by $\gamma_{ij}^*$ and $\gamma^*$, and assume for
simplicity that $\gamma_{ij}^* = \gamma_i^* \delta_{ij}$. (As above, the
general case is solved by finding the eigenvalues an eigenvectors of
$\gamma_{ij}^*$.) The solutions of the renormalization group equations in the
neighborhood of the fixed point become
\eqn\fixedsoln{\eqalign{
&{ c_i \left( \mu_1 \right) \over c_i \left( \mu_2 \right) }
= \left[ {\mu_1 \over \mu_2 }\right]^{\gamma_i^*} ,\cr
& \exp\left[-\int_{\mu_2}^{\mu_1} \gamma_\phi \left( \mu \right) d \log \mu
\right]
= \left[ {  \mu_1 \over  \mu_2 }\right]^{-\gamma^*}.
}}
so that \finalrg\ becomes
\eqn\gfixed{
G_n\left( \left\{s x \right\} ,c_{i}\left(\Lambda \right),\Lambda \right)=
s^{n\left(2-D\right)/2} s^{n \gamma^*}
G_n\left(\left\{ x \right\} ,s^{d_{i}-\gamma_i^*}
c_{i}\left(\Lambda/s\right),\Lambda\right)
}
This equation shows that scale invariance is recovered in the quantum theory
at an infrared stable fixed point, but the quantum dimensions of fields and
operators differ from their classical values. Operators now have dimension
$D-d_i+\gamma_i^*$, their coefficients have dimension $d_i-\gamma_i^*$, and
fields have dimension $\left(D-2\right)/2 - \gamma^*$. This is the reason
why $\gamma, \gamma_{ij}$ are called anomalous dimensions.  The classification
into relevant, irrelevant and marginal operators is the same as before, except
that one should use the quantum dimension of the operator which includes the
anomalous dimension.

\begfig 0 cm
\centerline{\hbox{
\epsfxsize=6cm\epsffile{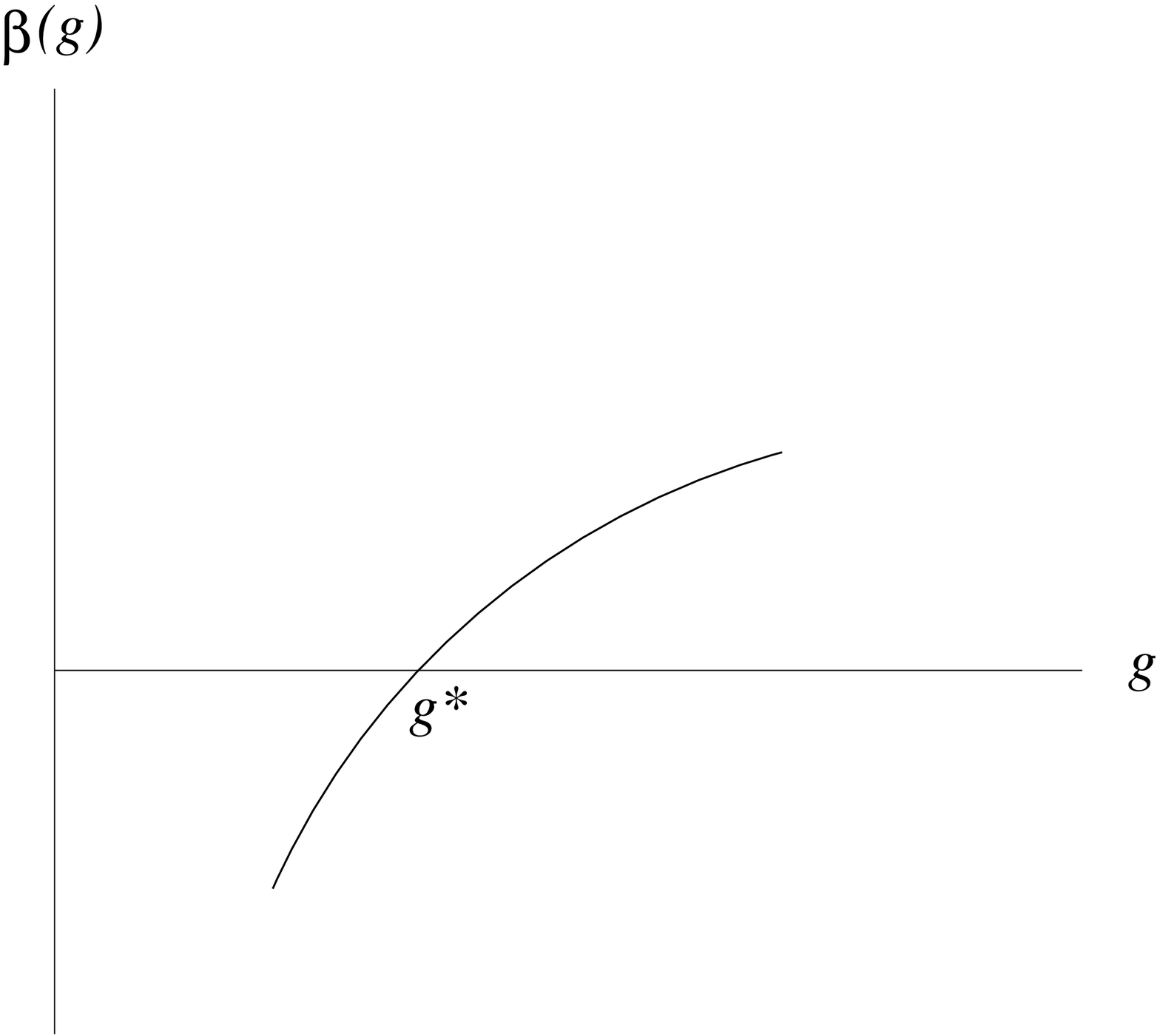}}}
\figure{\zerobeta}{An infrared stable fixed point of the $\beta$ function}
\endfig

In weakly coupled theories, operator anomalous dimensions can be computed in
perturbation theory, and are small. Thus quantum corrections cannot affect
which operators are relevant or irrelevant, since the classical dimensions of
operators are restricted to be integers or half-integers. The only effect of
quantum corrections is to turn marginal operators into relevant or irrelevant
operators, depending on whether their anomalous dimension is negative or
positive. In strongly coupled theories, more interesting effects can occur.
For example, in walking technicolor theories it is believed that a composite
operator $\bar \psi \psi$ with classical dimension 3 behaves in the quantum
theory as a scalar field with dimension 1, i.e. $\bar \psi \psi$ has anomalous
dimension $-2$. A solvable example of this kind exists in two dimensions.
The
two dimensional Thirring model with a fundamental fermion field
\eqn\thlag{
\CL = \bar \psi \left( i \delsl - m \right) \psi - {1 \over 2}
g \left(\bar \psi \gamma^\mu
\psi \right)^2,
}
is dual to the sine-Gordon model with a fundamental scalar field
\eqn\sGlag{
\CL = {1 \over 2} \partial_\mu \phi \partial^\mu \phi + {\alpha\over\beta^2}
\cos \beta\phi,
}
where the coupling constants $g$ and $\beta$ are related by
\eqn\tsg{
{\beta^2 \over 4 \pi} = {1 \over 1 + g/\pi}.
}
The fermion of the Thirring model is the sine-Gordon soliton, and the boson of
the sine-Gordon model is a fermion-antifermion bound state in the Thirring
model. The mapping \tsg\ shows that the strongly coupled sine-Gordon model with
$\beta^2\approx 4\pi$ can be mapped onto a weakly coupled Thirring model with
$g \approx 0$. There are two alternate descriptions of the same theory: (a) A
strongly interacting boson theory with large anomalous dimensions\fonote{For
example, the operator $\cos \beta \phi$ gets mapped to the Fermion mass term
$\bar \psi \psi$. In two dimensions, the canonical dimensions of
$\cos \beta \phi$ and $\bar \psi \psi$ are zero and one, respectively.}
(b) A weakly interacting fermion theory with small anomalous dimensions.
Formally, both descriptions are identical, but clearly (b) is better for
doing practical calculations.

The scaling dimensions of fields was determined from the free Lagrangian.
That is because one assumes that the effective Lagrangian can be written
as a weakly coupled field theory in terms of correctly chosen degrees of
freedom at low energies. If the degrees of freedom are strongly coupled,
the scaling dimension of the fields may change from their canonical value, as
we saw in the sine-Gordon model at $\beta^2=4\pi$.
Often, the most difficult task in
writing down an effective theory is {\it choosing the right degrees of
freedom}. In the sine-Gordon model, it is better to use a weakly coupled
soliton field $\psi$ instead of the fundamental field $\phi$ if $\beta^2
\approx 4 \pi$,
i.e. the effective Lagrangian for the sine-Gordon model with $\beta^2
\approx 4 \pi$ is the
Thirring  model. Low energy QCD is a weakly coupled
theory when written in terms of pion fields, but not when written in
terms of quark and gluon fields.\foot{This is obvious in the large $N_c$
limit, where one has a weakly interacting theory of mesons and baryons, with
a coupling constant $1/N_c$.}\ At low energies, the Goldstone boson
fields scale with canonical dimension zero (they are like angles), which
is different from $\bar q q$, which has dimension $3$ in free field theory.
There are many examples of this kind in condensed matter
physics. For example, in Landau Fermi liquid theory, the degrees of freedom
are weakly interacting quasiparticles, not the strongly interacting
electrons.

\titleb{}{SUMMARY}
We can now summarize the results of this section.
\medskip
\item{1.} Find a good set of variables to describe the dynamics.
\item{2.} Write down the effective action as a sum of operators, $ \sum_i c_i
\CO_i $.
\item{3.} The scaling rule is that $c_i \rightarrow s^{d_i-\gamma_i} c_i$,
where $d_i$ is the naive dimension and $\gamma_i$ is the anomalous dimension.
The most important operators are those of lowest dimension. Hopefully, a good
choice has been made in (1), so that the anomalous dimensions are small.
\item{4.}  To include all corrections up to order $1/s^r$, one should include
all operators with dimension $\le D+r$, i.e. all terms with coefficients of
dimension $\ge -r$.
\medskip
\noindent There are a finite number of operators that contribute to a given
order in $1/s$. In four dimensions, the dimensions of scalar, spinor and
vector fields is
\eqn\fdim{
\left[ \phi \right] = 1,\qquad
\left[ \psi \right] = 3/2,\qquad
\left[ A_\mu \right] = 1.
}
The allowed Lorentz invariant and gauge invariant operators of dimension $\le
4$ are $\phi^n, n \le 4$, $\bar \psi \psi$, $\partial_\mu \phi
\partial^\mu \phi$, $\bar \psi \dsl \psi$, $\bar \psi \psi \phi$, $F_{\mu \nu}
F^{\mu \nu}$.

\titlea{3}{Renormalizable Theories vs Effective Theories}
Field theory textbooks argue that a quantum field theory should be
renormalizable, i.e. that
the Lagrangian contain only terms with dimension $\le
D$. Otherwise one needs an infinite number of counterterms, hence an infinite
number of unknown parameters, and the theory has no predictive power.

An effective field theory Lagrangian contains an infinite number of terms. Let
us write the Lagrangian in the form
\eqn\lagexp{
\CL_{\rm eft} = \CL_{\le D} + \CL_{D+1}+ \CL_{D+2} + \ldots,
}
where $\CL_{\le D}$ contains all terms with dimension $\le D$, $\CL_{D+1}$
contains terms with dimension $D+1$, $\CL_{D+2}$ contains terms with dimension
$D+2$, and so on. The usual renormalizable Lagrangian is just the first term,
$\CL_{\le D}$. There are an infinite number of terms in $\CL_{\rm eft}$,
but one still has {\it approximate predictive power}. The effective Lagrangian
is used to compute processes at some scale $\Lambda/s$, where $\Lambda$ is the
scale of (possibly unknown) high energy interactions. One can compute with an
error of $1/s$ by retaining only $\CL_{\le D}$. Furthermore, one can extend
the approximation in a systematic way -- to compute with an error of order
$1/s^{r+1}$, one needs to retain terms up to $\CL_{D+r}$. There are only a
finite number of parameters to compute to a given order in $1/s$, so the
theory has predictive power.
\medskip
{\bf A non-renormalizable theory is just as good as a renormalizable theory
for computations, provided one is satisfied with a finite accuracy.}
\medskip
The usual renormalizable field theory result is recovered if one takes the
separation of scales $s \rightarrow \infty$. In this case, one can compute
using a renormalizable Lagrangian $\CL_{\le D}$ with no errors. While exact
computations are nice, they are irrelevant. Nobody knows
the exact theory up to infinitely high energies.
Thus any realistic calculation is done using an effective field theory. The
standard ``exact'' textbook analysis of QED is really an approximate
calculation in which terms suppressed by powers of $1/s$ have been neglected.

\titlea{4}{Two Simple Examples}
We now consider two simple examples that illustrates the utility of the
effective field theory method.

\titleb{}{Rayleigh Scattering}
The first example is Rayleigh scattering, the scattering of photons off atoms
at low energies.
Here low energies means energies small enough that one does not excite the
internal
states of the atom, or cause it to ionize. The atom can be treated as a
particle of mass $M$, interacting with the electromagnetic field. Let
$\psi(x)$
denote a field operator that creates an atom at the point $x$. Then the
effective Lagrangian for the atom is
\eqn\Ii{
\CL = \psi^\dagger\left(i \partial_t - {p^2\over 2M} \right) \psi + \CL_{\rm
int},
}
where $\CL_{\rm int}$ is the interaction term. Since the atom is neutral, the
interaction term is a function of the electromagnetic field strength
$F_{\mu\nu}=({\bf E},{\bf B})$. Gauge invariance forbids terms which
depend only on the
vector potential $A_\mu$. At low energies, the dominant interaction is one
which involves the smallest number of derivatives, and the smallest number of
photon fields, and has the form
\eqn\Iii{
\CL_{\rm int} = a_0^3 \ \psi^\dagger \psi \left(c_1 E^2 +c_2 B^2\right)
}
The electromagnetic field strength has mass dimension two, $\psi$ has mass
dimension 3/2 ($\psi^\dagger i \partial_t \psi$ has dimension four),
so that $c_1 a_0^3$ and $c_2 a_0^3$ have mass dimension $-3$. The typical
momentum scale is set
by the size of the atom $a_0$, so one expects $c_{1,2}$ to be
of order unity. The interaction \Iii\ gives the scattering amplitude $\CA \sim
c_i a_0^3 \omega^2$, since the electric and magnetic fields are
gradients of the vector potential, so each factor of $\bf E$ or $\bf B$
produces a factor of $\omega$. The scattering cross-section is proportional
to $\abs{c_i}^2 a_0^6 \omega^4$. This has the correct dimensions to be a
cross-section, so the phase-space is dimensionless, and one finds that
\eqn\Iiii{
\sigma \propto a_0^6\ \omega^4.
}
This reproduces the well-know $\omega^4$ dependence of the Rayleigh scattering
cross-section, which explains why the sky is blue. One can actually do
better, and determine the factors of $4\pi$ in \Iiii, but I won't discuss
that here. Equation \Iiii\ has corrections of order $\omega/a_0$ from higher
dimension operators which have been neglected in \Iii.

\titleb{}{The Euler-Heisenberg Lagrangian}
The Euler-Heisenberg effective Lagrangian is the effective Lagrangian for
photon-photon scattering at energies much lower than the electron mass $m_e$.
The leading order Lagrangian is the free Maxwell theory,
\eqn\maxwell{
\CL = -{1\over 4} F_{\mu \nu}F^{\mu \nu}.
}
The first interactions that can occur are from higher dimension operators. The
lowest non-trivial operators must contain four factors of the field strength
$F_{\mu\nu}$ and hence must be of dimension eight,
\eqn\eh{
\CL = {\alpha^2 \over m_e^4} \left[ c_1
\left(F_{\mu \nu}F^{\mu \nu}\right)^2 +  c_2
\left(F_{\mu \nu} \tilde F^{\mu \nu}\right)^2 \right].
}
(Terms with only three field strengths are forbidden by charge conjugation
symmetry.) The effective interaction \eh\ is generated from the box diagram of
Fig.~\fig\eulerfig{Light by light scattering in (a) QED and (b) in the
Euler-Heisenberg effective theory}.
The box diagram contains four factors of the electric charge $e$, and one
factor of $1/16\pi^2$ for the loop. In addition, the only dimensionful
parameter other than the external momenta is the electron mass $m_e$. This
allows us to write the Lagrangian in the form \eh, where $c_{1,2}$ are
dimensionless constants. An explicit computation gives
\eqn\cvalues{
c_1 = {1 \over 90}, \qquad c_2= {7 \over 90}.
}
The low energy cross-section for $\gamma\gamma \rightarrow \gamma \gamma$ is
obtained from the graph in the effective theory, Fig.~\eulerfig(b). The
scattering
amplitude is $\CA \sim \alpha^2 \omega^4/m_e^4$, since each gradient of the
photon field in \eh\ produces one factor of $\omega$. This produces a
cross-section of order
\eqn\ggsigma{
\sigma \sim \left( {\alpha^2 \omega^4 \over m_e^4 } \right)^2
{1 \over \omega^2}.
}
The phase space factor $1/\omega^2$ is obtained using dimensional analysis.
The cross-section must have dimensions of area, so the phase space must have
dimension $-2$. The only dimensionful parameter in the effective theory is the
photon energy $\omega$, so the phase space must be proportional to
$1/\omega^2$. Thus we find $\sigma \sim \alpha^4 \omega^6/m_e^8$, with an error
of order $\omega^2/m_e^2$ from neglected higher order interactions in \eh.

\begfig 0 cm
\centerline{\hbox{
\epsfxsize=8cm\epsffile{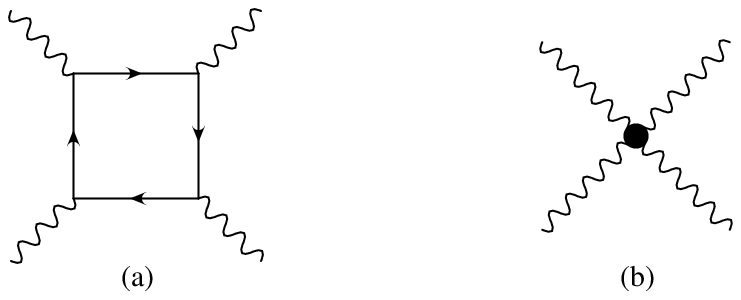}}}
\figure{\eulerfig}{Light by light scattering in (a) QED and (b) in the
Euler-Heisenberg effective theory. The solid dot represents the four-photon
interaction from the effective Lagrangian \eh}
\endfig

\titlea{5}{Weak Interactions at Low Energies: Tree Level}
The classic example of an effective field theory is the Fermi theory of weak
interactions. We first discuss how to obtain the Fermi theory as the
low-energy
limit of the renormalizable $SU(2)\times U(1)$ electroweak theory at tree
level. The use of effective field theories for the tree level weak
interactions
will seem at first like applying a lot of unnecessary formalism to a trivial
problem; the usefulness of the effective field theory method will only become
apparent after we study the $\Delta S=2$ weak interactions, which involve loop
corrections in field theory. Finally, we will discuss the weak interactions
including the leading logarithmic QCD corrections, for which the effective
field theory method is indispensable.

The basic flavor changing vertex in the quark sector is
the $W$ coupling to the quark current
\eqn\vertex{
-{ig\over\sqrt2}\ V_{ij}\ \bar q_i\, \gamma^\mu\, P_L\, q_j, }
where $V_{ij}$
is the Kobayashi-Maskawa mixing matrix, and $P_L=(1-\gamma_5)/2$ is the
left-handed projection operator.  The lowest order $\Delta S=1$ amplitude
arises from single $W$ exchange (Fig.~\fig\figwexch{$W$ exchange diagram for
the $\Delta S=1$ weak interactions}),
\eqn\wexch{
\CA = \left({ig\over\sqrt2}\right)^2 V_{us} V_{ud}^*\
\left(\bar u\, \gamma^\mu\, P_L\, s\right)
\left(\bar d\, \gamma^\nu\, P_L\, u\right)
\left({-ig_{\mu\nu}\over p^2-M_W^2}\right),}
where the $W$ boson propagator is in 't~Hooft-Feynman gauge, $p$ is the
momentum transferred by the $W$, and $u$, $d$, $s$ are quark spinors. The
exchange of unphysical scalars $\phi^{\pm}$ can be neglected, since their
Yukawa couplings to the light quarks are very small. The amplitude \wexch\
produces a non-local four-quark interaction, because of the factor of
$p^2-M_W^2$ in the denominator. However, if the momentum transfer $p$ is small
compared with $M_W$, the non-local interaction can be approximated by a local
interaction using the Taylor series expansion
\eqn\taylor{
{1\over p^2-M_W^2} = -{1\over M_W^2}\left(1+{p^2\over M_W^2} + {p^4\over
M_W^4}
+ \ldots\right), }
and retaining only a {\it finite} number of terms. To lowest order, the
amplitude is
\eqn\wexchz{
\CA = \bfrac i{M_W^2}\left({ig\over\sqrt2}\right)^2 V_{us} V_{ud}^*\
\left(\bar u\, \gamma^\mu\, P_L\, s\right)\left(\bar d\, \gamma_\mu\, P_L\, u
\right)+\CO\left({1\over M_W^4}\right).
} The amplitude \wexchz\ can be obtained using the effective Lagrangian
\eqn\lexchz{
\CL = -{4 G_F\over\sqrt2}V_{us} V_{ud}^*\
\left(\bar u\, \gamma^\mu\, P_L\, s\right)\left(\bar d\, \gamma_\mu\, P_L\, u
\right)+\CO\left({1\over M_W^4}\right),
} where $u$, $d$ and $s$ are now the quark fields, and we have used the
definition
\eqn\gfermidef{
{G_F\over\sqrt2} \equiv {g^2\over 8 M_W^2}.
}
The effective Lagrangian \lexchz\ can be used to study the weak decays of
quarks at low energies. The basic interaction is a local four-Fermion vertex,
as shown in Fig.~\fig\figtwo{The effective four-Fermi interaction of \lexchz.
This
interaction reproduces the results of \figwexch\ to order $1/M_W^2$}.
To avoid complications with hadronic
matrix elements and QCD corrections (which will be discussed later), consider
instead the effective Lagrangian for $\mu$ decay
\eqn\muleff{
\CL = -{4 G_F\over\sqrt2}\
\left(\bar e\, \gamma^\mu\, P_L\, \nu_e\right)\left(\bar \nu_\mu\,
\gamma^\mu\,
P_L\, \mu\right) +\CO\left({1\over M_W^4}\right), } whose derivation is almost
identical to that of \lexchz.  Using \muleff, neglecting the $1/M_W^4$
terms, and integrating over phase space gives the standard result for the muon
lifetime at lowest order,
\eqn\mulife{
\Gamma_\mu = {G_F^2 m_\mu^5\over 192\pi^3}.
} This calculation is well known, and will not be repeated here.
\bigskip\noindent
{\it To summarize:} at lowest order, the ``full theory,'' which is the $SU(2)
\times U(1)$ electroweak theory, can be replaced by the ``effective theory,''
which is QED plus the effective Lagrangian \lexchz\ (or \muleff), up
to
corrections of order $1/M_W^4$. The effective theory can be used to compute
physical processes such as the muon lifetime.  So far, the effective field
theory method is a fancy way of saying that we have approximated the $W$ boson
propagator in Fig.~\figwexch\ by $1/M_W^2$. The real advantage of the effective
field theory method will be apparent after we have discussed the one-loop
$\Delta S=2$ amplitude including QCD radiative corrections.

\begfig 0 cm
\centerline{\hbox{
\epsfxsize=3.5cm\epsffile{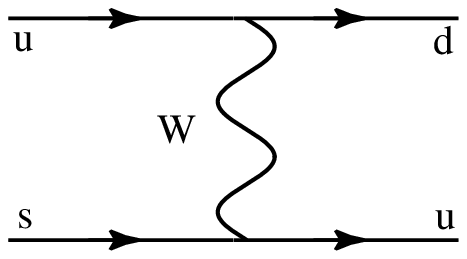}}}
\figure{\figwexch}{$W$ exchange diagram for the $\Delta S=1$ weak interactions}
\endfig

\begfig 0 cm
\centerline{\hbox{
\epsfxsize=3.5cm\epsffile{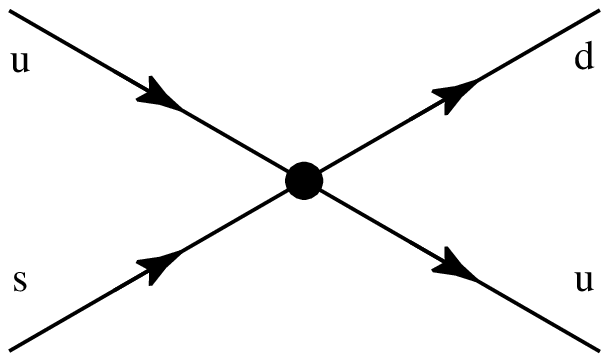}}}
\figure{\figtwo}{The effective four-Fermi interaction of \lexchz. This
interaction reproduces the results of Fig.~\figwexch\ to order $1/M_W^2$}
\endfig

\titlea{6}{Renormalization in Effective Field Theories}
In quantum field theory, knowing the Lagrangian is not sufficient to compute
results for physical quantities. In addition, one needs to specify a way to
get
finite, unambiguous answers for physical quantities. In perturbation theory,
this corresponds to a choice of renormalization scheme which (i) regulates the
integrals and (ii) subtracts the infinities in a systematic way.  The
effective
Lagrangian \lexchz\ that we have constructed is non-renormalizable, since
it contains an operator of dimension six, times a coefficient $G_F$ which is
of
order $1/M_W^2$. The neglected $1/M_W^4$ term contains operators of dimension
eight, and so on. To use the effective Lagrangian beyond tree level, it is
necessary to give a renormalization scheme as part of the definition of the
effective field theory.  Without this additional information, the effective
Lagrangian \lexchz\ is meaningless.

It is important to keep in mind that the effective field theory is a {\it
different theory} from the full theory. The full theory of the weak
interactions is a renormalizable field theory. The effective field theory is a
non-renormalizable field theory, and has a different divergence structure from
the full theory. The effective field theory is constructed to correctly
reproduce the low-energy effects of the full theory to a given order in
$1/M_W$. The effective Lagrangian includes more terms as one works to higher
orders in $1/M_W$. The effective field theory method is useful only for
computing results to a certain order in $1/M_W$.  If one is interested in the
answer to all orders in $1/M_W$, it is obviously much simpler to use the full
theory.

The renormalization scheme must be carefully chosen to give a sensible
effective field theory. To see what the possible problems might be, consider
the flavor diagonal effective Lagrangian from $W$ and $Z$ exchange
\eqn\diageff{
\CL = -{4 G_F\over\sqrt2}V_{ui} V_{ui}^*\
\left(\bar u\, \gamma^\mu\, P_L\, q_i\right)\left(\bar q_i\, \gamma_\mu\,
P_L\,
u\right) + \left({\rm Z-exchange}\right) + \CO\left({1\over M_W^4}\right), }
where $i=d,s,b$. At tree level, the $W$ and $Z$ exchange graphs contribute to
flavor diagonal parity violating $u$-quark interactions at order
$G_F\sim1/M_W^2$.  At one loop, the interaction \diageff\ induces a $Z\bar
u u$ vertex from the graph in Fig.~\fig\figzloop{Z loop}\ which is of the form
\eqn\intone{
I\sim{1\over M_W^2} \int d^4 k\ {1\over k^2}, } neglecting the $\gamma$-matrix
structure. The $1/k^2$ factor is from the two fermion propagators in the loop,
and $G_F$ has been rewritten as $G_F\sim1/M_W^2$.  Since the effective field
theory is valid up to energies of order $M_W$, one can estimate the integral
using a momentum space cutoff $\Lambda$ of order $M_W$,
\eqn\inttwo{
I \sim {1\over M_W^2} \Lambda^2 \sim \CO\left(1\right).  } Thus the
interaction \diageff\ produces a one loop correction
to the $Z\bar u u$ vertex of
order
one.  Similarly, one can show that higher order terms, such as the dimension
eight operators, are all equally important. A loop graph of the form
Fig.~\figzloop\ (where the vertex is now a dimension eight operator) is of
order
\eqn\intthree{
I^\prime\sim {1\over M_W^4} \int d^4k\ {1\over k^2} k^2 \sim {\Lambda^4\over
M_W^4} \sim \CO\left(1\right), } etc. The additional $k^2$ in the integral
\intthree\ is from the extra $\partial^2$ at the four-quark vertex in the
dimension eight operator arising from the order $p^2$ term in the expansion of
\lexchz. The loop graph with an insertion of the dimension eight operator
is just as important as the loop graph with an insertion of the dimension six
operator; both are of order unity and cannot be neglected. Similarly, all the
higher order terms in the effective Lagrangian are equally important, and the
entire expansion breaks down. A similar problem also occurs in the flavor
changing $\Delta S=1$ weak interactions that we have been studying, but the
analysis is more subtle because of the GIM mechanism, which is why we
considered the $Z\bar u u$ vertex.

\begfig 0 cm
\centerline{\hbox{
\epsfxsize=4cm\epsffile{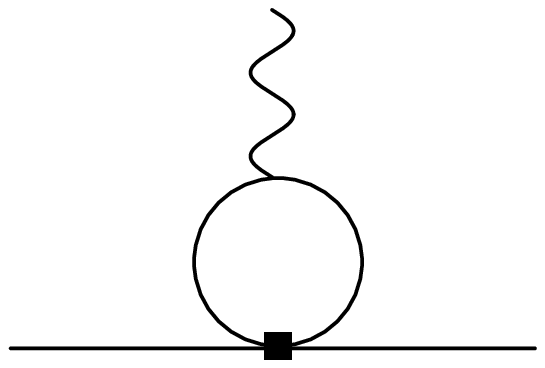}}}
\figure{\figzloop}{One loop correction to the $Z\bar u u$ vertex. The
solid square represents either the dimension six four-quark interaction of
eq.~\diageff, or the dimension eight four-quark operator discussed in the text}
\endfig

The effective field theory expansion breaks down if one introduces a
mass-dependent subtraction scheme such as a momentum space cutoff.\fonote{One
way to solve this problem is to use a cutoff $\Lambda \ll M_W$. This method
does not allow one to easily match between the full and effective theories, or
to include QCD corrections.}\ This problem
can be cured if one uses a mass-independent subtraction scheme, such as
dimensional regularization and minimal subtraction, in which the dimensional
parameter $\mu$ only appears in logarithms, and never as explicit powers such
as $\mu^2$. In such a subtraction scheme $\beta$-functions and anomalous
dimensions of composite operators are mass independent.  If one estimates the
integrals \intone\ and \intthree\ in a mass-independent subtraction
scheme, one finds
\eqn\inti{\eqalign{
I=&{1\over M_W^2} \int d^4 k\ {1\over k^2} \sim {m^2\over M_W^2}\log \mu,\cr
I^\prime=&{1\over M_W^4} \int d^4 k\ {1\over k^2}k^2 \sim {m^4\over
M_W^4}\log\mu, }} where $m$ is some dimensionful parameter that is {\it not}
the renormalization scale $\mu$. It must be some other dimensionful scale that
enters the loop graph of Fig.~\figzloop, such as the quark mass or the external
momentum. This completely changes the estimate of the integrals. The integrals
are no longer of order one, but are small provided $m\ll M_W$.  As a result:
\medskip
\item{1.} The effective Lagrangian produces a well-defined expansion of the
weak amplitudes in powers of $m/M_W$, where $m$ is some low scale such as the
quark mass or the external momentum (or $\lqcd$ when one includes QCD
effects).
One has a systematic expansion in powers of some low scale over $M_W$. This
makes precise what is meant by neglecting $1/M_W^4$ terms in \wexchz.
\item{2.} Loop integrals do not have a power law dependence on $\mu\sim M_W$,
so one can count powers of $1/M_W$ directly from the effective
Lagrangian. Graphs with one insertion of terms in $\leff$ of order $1/M_W^2$
produces amplitudes of order $1/M_W^2$. Graphs with one insertion of terms of
order $1/M_W^4$ or two insertions of terms of order $1/M_W^2$ produce
amplitudes of order $1/M_W^4$, etc.
\item{3.} The effective field theory behaves for all practical purposes like
a renormalizable field theory if one works to some fixed order in $1/M_W$.
This is because there are only a finite number of terms in $\leff$ that are
allowed to a given order in $1/M_W$. Terms of higher order in $1/M_W$ can be
safely neglected because they can never be multiplied by positive powers of
$M_W$ to produce effects comparable to lower order terms.
\medskip
\noindent It is well-known that different renormalization schemes lead to
equivalent answers for all physical quantities. In an effective field theory,
a
mass-independent subtraction scheme is particularly convenient, since it
provides an efficient way of keeping only a few operators in $\leff$, and in
deciding which Feynman graphs are important. Nevertheless, one must be able to
obtain the same results in a mass-dependent scheme such as a momentum space
cutoff. This is true in principle: a mass dependent scheme has an infinite
number of contributions that are of leading order (from the dimension four,
six, eight, $\ldots$, operators). If one resums this contribution, then the
remaining effects (again from an infinite number of terms) will be of order
$1/M_W^2$. Resumming the latter leaves a contribution of $1/M_W^4$, etc.  The
net result of this procedure is to reproduce the same answer as that obtained
much more simply using a mass-independent renormalization scheme.  The
connection between different renormalization schemes is much more complicated
in an effective field theory (which is non-renormalizable), than in a
renormalizable field theory.

\titlea{7}{Decoupling of Heavy Particles}
There is one important drawback to using a mass-independent subtraction
scheme -- heavy particles do not decouple.
\fonote{A mass independent subtraction
scheme does not satisfy the conditions for the Appelquist-Carazzone theorem.}
\ This must obviously be true since the contribution of
particles to $\beta$-functions does not depend on the particle mass. For
example, a 1~TeV charged lepton makes the same contribution as an electron to
the QED $\beta$-function at 1~GeV.

It is instructive to look at the contribution of a charged fermion to the
$\beta$-function in QED. Evaluating the diagram of Fig.~\fig\figqed{qed} in
dimensional regularization gives
\eqn\qedi{
i{e^2\over 2\pi^2}\left(p_\mu p_\nu -p^2 g_{\mu\nu}\right)
\left[{1\over 6\epsilon} - {\gamma\over 6} - \int_0^1 dx\ x(1-x)\
\log{m^2-p^2 x(1-x)\over 4\pi\mu^2}\right],
} where $p$ is the external momentum, $m$ is the fermion mass, $\gamma$ is
Euler's constant, and $\mu$ is the scale parameter of dimensional
regularization.

\begfig 0 cm
\centerline{\hbox{
\epsfxsize=5cm\epsffile{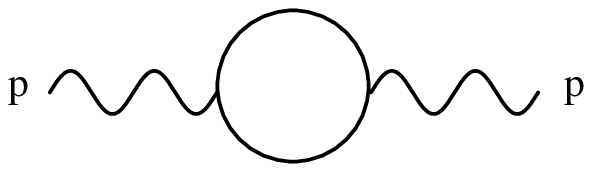}}}
\figure{\figqed}{One loop contribution to the QED $\beta$-function
from a fermion of mass $m$}
\endfig

\titleb{}{Mass-Dependent Scheme}
In a mass-dependent scheme, such as an off-shell momentum space subtraction
scheme, one subtracts the value of the graph at a Euclidean momentum point
$p^2=-M^2$, to get
\eqn\qediv{
-i{e^2\over 2\pi^2}\left(p_\mu p_\nu - p^2 g_{\mu\nu}\right)
\left[\int_0^1 dx\ x(1-x)\ \log{m^2-p^2 x(1-x)\over m^2+M^2 x(1-x)}\right].
} The fermion contribution to the QED $\beta$-function is obtained by acting
with $(e/2)M d/dM$ on the coefficient of $i\left(p_\mu p_\nu -p^2
g_{\mu\nu}\right)$,
\eqn\qedv{\eqalign{
\beta\left(e\right)&=-\bfrac e 2 M{d\over dM} {e^2\over 2\pi^2}
\left[\int_0^1 dx\ x(1-x)\ \log{m^2-p^2 x(1-x)\over m^2+M^2 x(1-x)}\right]\cr
&= {e^3\over 2\pi^2}\int_0^1 dx\ x(1-x)\ {M^2 x (1-x)\over m^2+M^2 x(1-x)}.
}}
The fermion contribution to the $\beta$-function is plotted in
Fig.~\fig\figqedbeta{QED beta}. When the fermion mass $m$ is small compared
with the renormalization point $M$, $m\ll M$, the $\beta$-function
contribution
is
\eqn\betasmall{
\beta\left(e\right) \approx {e^3\over 2\pi^2}\int_0^1 dx\ x(1-x) = \bfrac
{e^3}
{12\pi^2}.  } As the renormalization point passes through $m$, the fermion
decouples, and for $M\ll m$, its contribution to $\beta$ vanishes as
\eqn\betalarge{
\beta\left(e\right) \approx \bfrac{e^3}{2\pi^2}
\int_0^1 dx\ x(1-x) {M^2 x (1-x)\over m^2}
=\bfrac {e^3}{60 \pi^2} \bfrac{M^2}{m^2}.  }

\begfig 0 cm
\centerline{\hbox{
\epsfxsize=6cm\epsffile{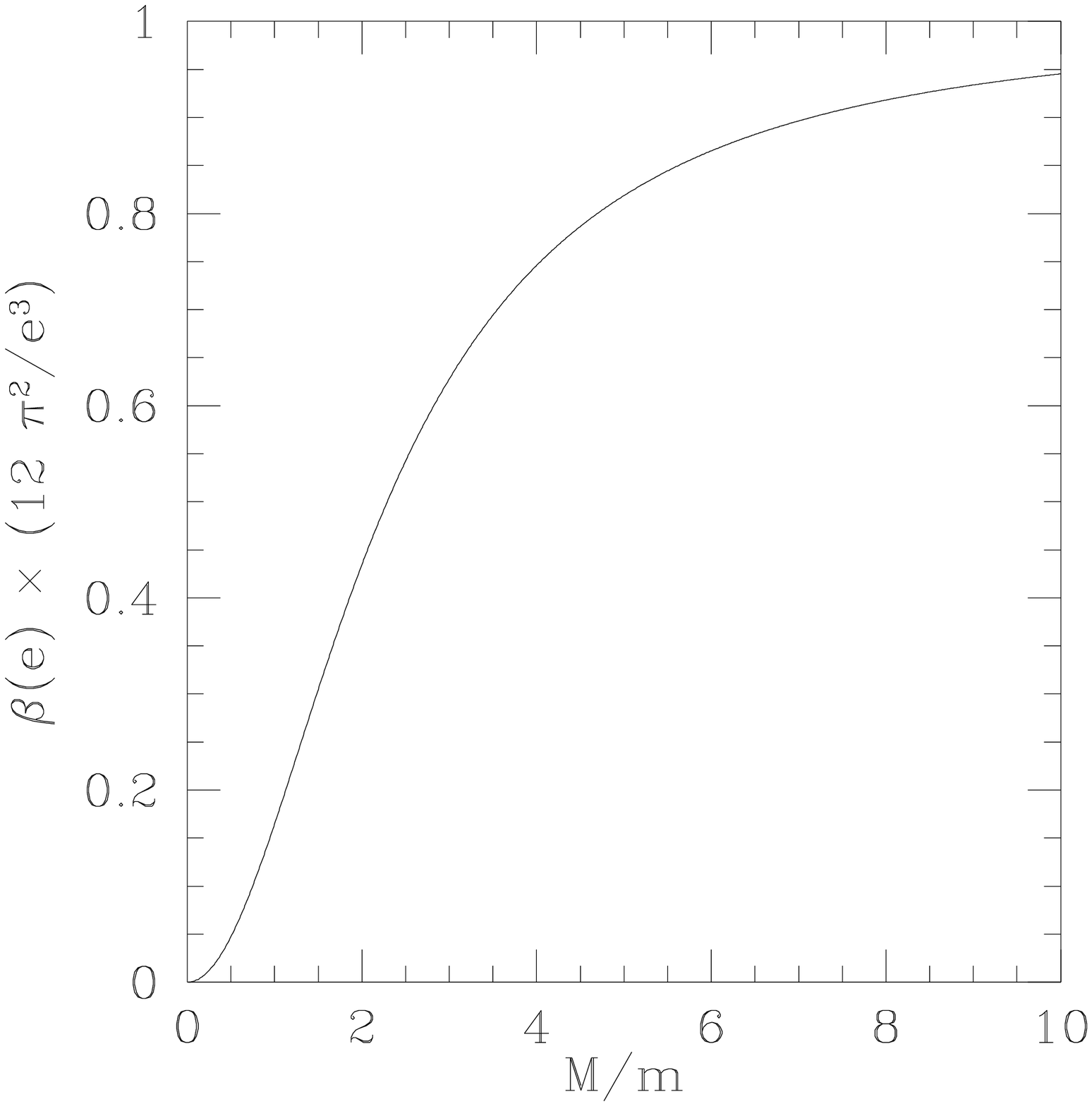}}}
\figure{\figqedbeta}{Contribution of a fermion of mass $m$ to the QED
$\beta$-function. The result is given for the momentum-space
subtraction scheme, with renormalization scale $M$. The $\beta$-function does
not attain its limiting value of $e^3/12\pi^2$ until $M \ga 10 m$.
The fermion decouples for $M\ll m$}
\endfig

\titleb{}{The $\bar{MS}$ Scheme}
In the $\bar {MS}$ scheme, one subtracts the $1/\epsilon$ pole and redefines
$4\pi\mu^2 e^{-\gamma}\rightarrow\mu^2$, to give
\eqn\qedii{
-i{e^2\over 2\pi^2}\left(p_\mu p_\nu - p^2g_{\mu\nu}\right)
\left[\int_0^1 dx\ x(1-x) \log{m^2-p^2 x(1-x)\over \mu^2}\right].
} The fermion contribution to the QED $\beta$-function is obtained by acting
with $(e/2)\mu d/d\mu$ on the coefficient of $i\left(p_\mu p_\nu -p^2
g_{\mu\nu}\right)$,
\eqn\qediii{\eqalign{
\beta\left(e\right)&=-\bfrac e 2\mu{d\over d\mu} {e^2\over 2\pi^2}
\left[\int_0^1 dx\ x(1-x) \log{m^2-p^2 x(1-x)\over \mu^2}\right]\cr
&= {e^3\over 2\pi^2}\int_0^1 dx\ x(1-x) = \bfrac {e^3}{12\pi^2}, }} which is
independent of the fermion mass and $\mu$.

The fermion contribution to the $\beta$-function in the $\bar{MS}$ scheme does
not vanish as $m\gg \mu$, so the fermion does not decouple as it should. There
is another problem: the finite part of the Feynman graph in the $\bar {MS}$
scheme at low momentum is
\eqn\qedfinite{
-i{e^2\over 2\pi^2}\left(p_\mu p_\nu - p^2g_{\mu\nu}\right)
\left[\int_0^1 dx\ x(1-x) \log{m^2\over \mu^2}\right],}
from \qedii.  For $\mu\ll m$ the logarithm becomes large, and perturbation
theory breaks down. These two problems are related.  The large finite parts
correct for the fact that the value of the running coupling used at low
energies is incorrect, because it was obtained using the ``wrong''
$\beta$-function.  The two problems can be solved at the same time by
integrating out heavy particles. One uses a theory including the fermion when
$m<\mu$, and a theory without the fermion when $m>\mu$. Effects of the heavy
particle in the low energy theory are included via higher dimension operators,
which are suppressed by inverse powers of the heavy particle mass.  The
matching condition of the two theories at the scale of the fermion mass is
that
$S$-matrix elements for light particle scattering in the low-energy theory
without the heavy particle must match those in the high-energy theory with the
heavy particle.

For the case of a spin-1/2 fermion at one loop, this implies that the running
coupling is continuous at $m=\mu$. The $\beta$-function is discontinuous at
$m=\mu$, since the fermion contributes $e^3/12\pi^2$ to $\beta$ above $m$ and
zero below $m$. The $\beta$-function is a step-function, instead of having a
smooth crossover between $e^3/12\pi^2$ and zero, as in the momentum-space
subtraction scheme. Decoupling of heavy particles is implemented by hand in
the
$\bar{MS}$ scheme by integrating out heavy particles at $\mu\sim m$. One
calculates using a sequence of effective field theories with fewer and fewer
particles. The main reason for using the $\bar{MS}$ scheme and integrating out
heavy particles is that it is much easier to use in practice than the
momentum-space subtraction scheme. Virtually all radiative corrections beyond
one-loop are evaluated in practice using the $\bar{MS}$ scheme.

There are some instances in which heavy particle effects are important in the
low energy effective theory. An example of this is the top quark in the
standard model. The reason is that the top quark has a mass $ m_t = g_t v/
\sqrt 2 $, where $g_t$ is the top quark Yukawa coupling, and $v$ is the
vacuum expectation value of the Higgs field. Taking $m_t$ large while keeping
$v$ fixed is equivalent to taking $g_t$ large. Diagrams involving top quarks
and scalars
(either the Higgs boson or the longitudinal parts of the $W$ and $Z$)
can be large, because they involve factors of $g_t$ which can cancel any
$1/m_t$ suppression. We will see an example of this in the next section,
where the $\Delta S = 2$ amplitude is shown to grow with $m_t$. One can still
integrate out the heavy top quark, but the low energy theory contains
operators
with coefficients which grow with $m_t$.

\titlea{8}{Weak Interactions at Low Energies: One Loop}
The ideas discussed so far can now be applied to the weak interactions at one
loop.  The amplitude for the $\Delta S=2$ amplitude for $K^0$-$\bar K^0$
mixing
is of order $G_F^2$. The leading contribution to this amplitude in the
standard
model is from the box diagram of Fig.~\fig\figbox{Box graph}, where one sums
over quarks
$i,j=u,c,t$ in the intermediate states. The sum of the $W$ and unphysical
scalar exchange graphs is
\def\abox{\CA^{\rm box}}
\eqn\inamilim{
\abox={g^4\over 128\pi^2 M_W^2}\ \sum_{i,j} \xi_i \xi_j\  \bar E(x_i,x_j)\
\left(\bar d\, \gamma^\mu\,P_L\,s\right)\left(\bar d\,
\gamma_\mu\,P_L\,s\right),
} where
\eqn\defxi{
x_i = {m^2_i\over M_W^2}, }
\eqn\xidef{
\xi_i = V_{is} V_{id}^*,
}
\eqn\ebarxy{\eqalign{
\bar E(x,y) &= - x y \Bigl\{ {1\over x-y}\left[ \bfrac 1 4 -
\bfrac 3 2 \bfrac 1 {x-1} - \bfrac 3 4 \bfrac 1 {(x-1)^2} \right]
\log x\cr & \qquad +  {1\over y-x}\left[ \bfrac 1 4 -
\bfrac 3 2 \bfrac 1 {y-1} - \bfrac 3 4 \bfrac 1 {(y-1)^2} \right]
\log y- \bfrac 3 4 \bfrac 1 {(x-1)(y-1)} \Bigr\},
}} and
\eqn\ebarxx{
\bar E(x,x) = - \bfrac 3 2 \left( \bfrac x {x-1}\right)^3 \log x
- x \left[ \bfrac 1 4 - \bfrac 9 4 \bfrac 1 {x-1} - \bfrac 3 2
\bfrac 1 {(x-1)^2}\right].
} In the limit $m_u=0$ and $m_{c,t}\ll M_W$,\fonote{The $\Delta S=2$ amplitude
is
considered in the limit $m_t\ll M_W$. This was the approximation used in the
original calculations, and makes it easier for the reader to compare with the
literature. It also simplifies the discussion somewhat, because the $t$-quark
and $c$-quark can be treated in a similar fashion.}
\eqn\inamilimit{
\abox=
-{G_F^2\over 4\pi^2}\ \left(\bar d\, \gamma^\mu\,P_L\,s\right)\left(\bar d\,
\gamma_\mu\,P_L\,s\right)
\left[ \xi_c^2\, {m_c^2} + \xi_t^2\, {m_t^2}
+ 2 \xi_c\xi_t\, m_c^2 \log \bfrac {m_t^2} {m_c^2}\right],
}
using \gfermidef.
The $\Delta S=2 $ amplitude
is of order $1/M_W^4$, rather than $1/M_W^2$ as one might
naively expect, because of the GIM mechanism: The quark mass independent piece
of the $\Delta S=2 $ amplitude is proportional to
\eqn\gim{
\xi_u+\xi_c+\xi_t = \sum_i V_{id} V_{is}^*=0,
} which vanishes because the KM matrix is unitary.

\begfig 0 cm
\centerline{\hbox{
\epsfxsize=5cm\epsffile{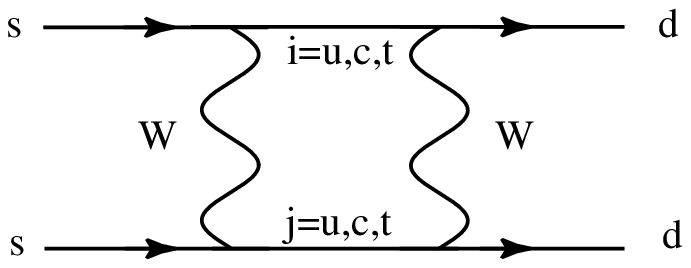}}}
\figure{\figbox}{The box diagram for the $\Delta S=2$
$K^0-\bar{K}^0$ mixing amplitude}
\endfig

\titleb{}{Matching at $M_W$}
We will now reproduce \inamilimit\ using an effective field theory
calculation to one loop. At the scale $M_W$, the $\Delta S=2$ amplitude in the
full theory is given by a loop graph in the effective theory involving two
insertions of the $\Delta S=1$ interaction, plus a local four-Fermi $\Delta
S=2$ interaction. The sum of the loop graph and the local $\Delta S=2$
interaction must reproduce the $\Delta S=2$ interaction in the full theory to
order $1/M_W^4$, as shown schematically in Fig.~\fig\figebox{figebox}.  The
tree
level graphs of Figs.~\figwexch\ and \figtwo\ are chosen to be the same in the
full
and effective theory to order $1/M_W^2$, but this does not imply that the loop
graphs in the full and effective theory are equal to order $1/M_W^4$. The two
loop graphs in Fig.~\figebox\ would be equal to order $1/M_W^4$ if the loop
graphs
in the full and effective theory were finite.  However, in general, the graphs
are infinite, and need subtractions. There is no simple relation between the
renormalization prescriptions in the full and effective theories and one needs
to add a local $\Delta S=2$ counterterm at the scale $M_W$, which is the
difference between the loop graphs in the full and effective theories.
The graphs in the effective theory are more divergent than in the full
theory. In our example, the box diagram in the full theory is convergent by
naive power counting.
\eqn\XViv{
I_{\rm full} \sim \int d^4k \left({1\over k}\right)^2
\left({1\over k^2}\right)^2,
}
whereas the graph in the effective theory is quadratically divergent,
\eqn\XVv{
I_{\rm eff} \sim \int d^4k \left({1\over k}\right)^2,
}
where we have used a factor of $1/k$ for each internal fermion line, and
$1/k^2$ for each internal boson line. In the case of the standard model, the
graph in the effective theory is more convergent than the naive estimate
because of the GIM mechanism. As we have seen, the fermion mass-independent
part of the diagram is proportional to $\xi_u + \xi_c +\xi_t$, which
vanishes. Thus the non-vanishing parts of the graphs in the full and effective
theory must involve a factor of the internal fermion mass. In fact, there have
to be two factors of the fermion mass because the $\Delta S=1$ vertex only
involves left-handed fields, and a fermion mass changes a left-handed fermion
to a right-handed fermion. Thus in the effective theory, the non-zero part of
the diagram must have two mass insertions on each of the fermion lines (there
is a separate GIM mechanism for each line because of the independent sums over
$i$ and $j$ in \inamilim), as represented in
Fig.~\figebox. This increases the degree of convergence of the diagram by
two for each internal quark line, and converts it from a diagram that diverges
like $k^2$ to a diagram that converges like $1/k^2$. Since the diagrams in the
full and effective theory are both finite, the local $\Delta S=2$ vertex
induced at the scale $M_W$ vanishes.

\begfig 0 cm
\centerline{\hbox{
\epsfxsize=12cm\epsffile{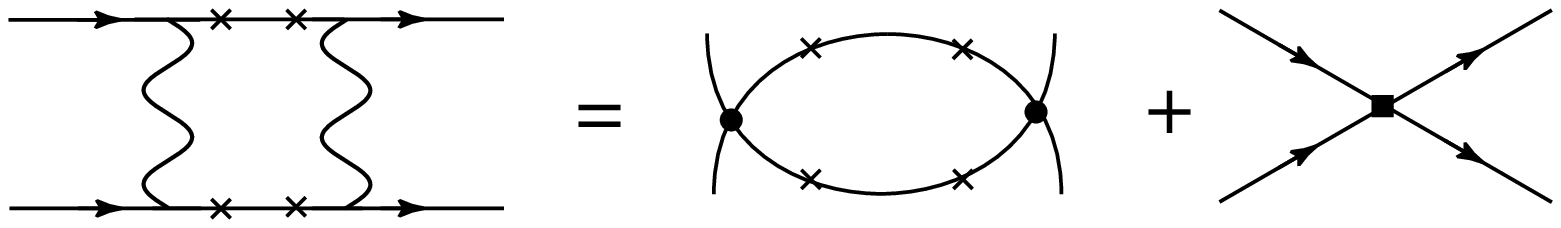}}}
\figure{\figebox}{Box diagram for the $\Delta S=2$ amplitude in
the full and
effective theories. The crosses represent fermion mass insertions. The solid
circle is a $\Delta S=1$ vertex, and the solid square is a local
$\Delta S=2$ vertex}
\endfig

\titleb{}{Matching at $m_t$}
The effective Lagrangian remains unchanged down to the scale $\mu=m_t$, if one
neglects QCD radiative corrections. At the scale $\mu=m_t$, one integrates out
the top quark. The ``full theory'' is now the effective Lagrangian including
six quarks, and the ``effective theory'' is the effective Lagrangian including
only five quarks. The $\Delta S=1$ interactions in the five-quark theory are
trivially obtained from those in the six-quark theory, by dropping all terms
that contain the $t$-quark. The $\Delta S=2$ interactions in the five- and
six-quark theories are given in
Fig.~\fig\figboxt{figboxt},
where the intermediate states in the six-quark theory are the $u$, $c$ and $t$
quarks, and in the five-quark theory are the $u$ and $c$ quarks. There is no
GIM cancellation once the top quark has been integrated out of the theory, so
the loop graph in the five-quark theory is divergent, and there will (in
principle) be a non-zero counterterm induced at the scale $m_t$.  The value of
the counterterm is the difference in the diagrams in the theories above and
below $m_t$, and so is given by the graphs in the theory above $m_t$ that
involve at least one $t$-quark in the loop, as shown in
Fig.~\fig\figtcterm{figtcterm}. All other graphs in the six-quark theory are
identical to the corresponding graphs in the five-quark theory. The loop
graphs
in the theory above $m_t$ can be calculated quite simply, and lead to the
matching condition
\eqn\matcht{
c_2(\mu=m_t-0)= {G_F^2\over 4\pi^2} \left[ \xi_t^2\, m_t^2 + 2 \xi_c\xi_t
\left(  m_t^2 +  m_c^2\right)+2 \xi_u\xi_t \left(m_t^2 +
m_u^2\right)\right], } where the contributions come from the finite part of
Fig.~\figtcterm, and $c_2$ is the coefficient of the $\Delta S=2$ operator
$\left(\bar d\, \gamma^\mu\,P_L\,s\right)\left(\bar d\,
\gamma_\mu\,P_L\,s\right)$. Using the
relation \gim\ and neglecting $m_u$ gives
\eqn\matchtII{
c_2(\mu=m_t-0)= {G_F^2\over 4\pi^2} \left[- \xi_t^2\, m_t^2 + 2 \xi_c\xi_t\,
m_c^2\right].  }

\begfig 0 cm
\centerline{\hbox{
\epsfxsize=12cm\epsffile{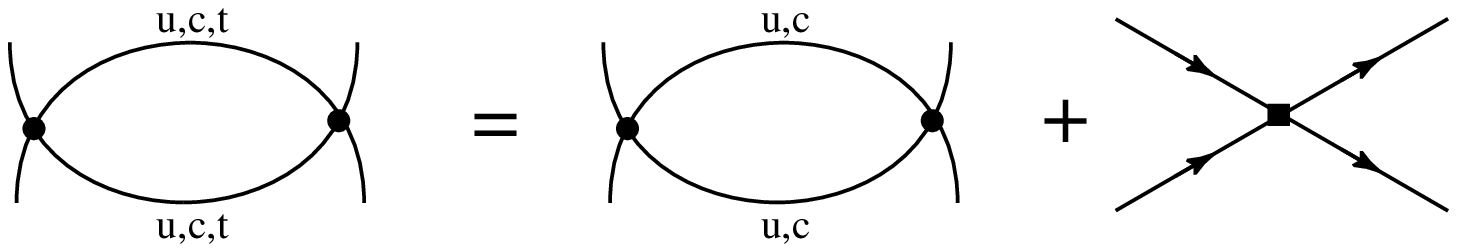}}}
\figure{\figboxt}{Matching condition at the $t$-quark scale. The
solid square is the $\Delta S=2$ counterterm induced at
$\mu=m_t$}
\endfig

\begfig 0 cm
\centerline{\hbox{
\epsfxsize=12cm\epsffile{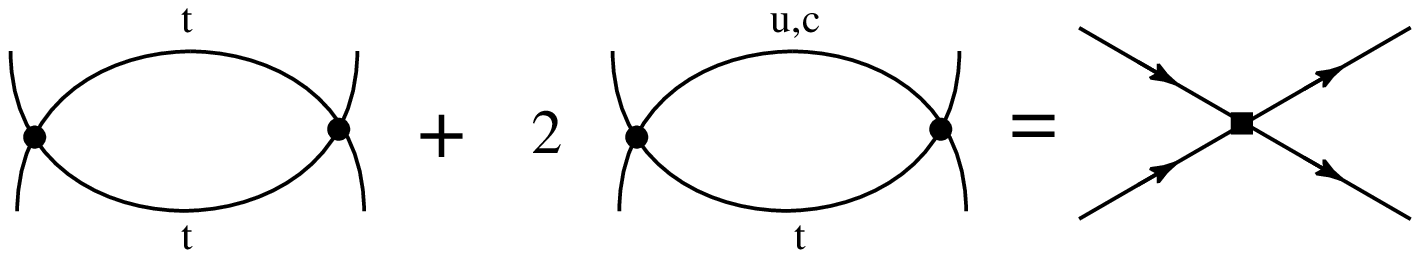}}}
\figure{\figtcterm}{Graphs to be computed to evaluate the $\Delta
S=2$ counterterm induced at $\mu=m_t$}
\endfig

Equation \matchtII\ is really the difference of two calculations at the scale
$m_t$ -- one in the full theory and one in the effective theory. Both
calculations are sensitive to infrared effects, such as confinement. However,
all infrared effects cancel in the difference, and
$c_2\left(\mu=m_t+0 \right)-c_2\left(\mu=m_t-0 \right)$ is not sensitive
to infrared effects.
An arbitrary infrared regulator can be used if the diagrams are infrared
divergent. The loop graphs will depend on the choice of regulator, but the
matching condition will not. The matching condition is only sensitive to
momenta of order $\mu=m_t$, so mass parameters such as $m_u$ are short
distance parameters such as the $\bar{MS}$ mass renormalized at $\mu=m_t$.

\titleb{}{Scaling from $m_t$ to $m_c$}
The next step is to scale from the scale $m_t$ to $m_c$. The loop graph
Fig.~\fig\figscale{figscale}\ is divergent, because there is no
longer a GIM mechanism in the five-quark theory, and $c_2$ is renormalized
proportional to $c_1^2$, where $c_1$ is the coefficient of the $\Delta S=1$
operator. This implies that there is a renormalization group equation for
$c_2$,
\eqn\rgct{
\mu {d\over d\mu} c_2 = \bfrac{1}{8\pi^2}\ {c_1^2\, m_c^2}\ \xi_c\xi_t,
} where the anomalous dimension is computed using the infinite part of
Fig.~\figscale. Integrating this equation from $m_t$ to $m_c$ gives
\eqn\rgint{\eqalign{
c_2(m_c) &= c_2(m_t) + \bfrac{1}{8\pi^2} {c_1^2 }
\ m_c^2\ \xi_c\xi_t\ \log\bfrac{m_c}{m_t},\cr
 &= c_2(m_t) - \bfrac {G_F^2}{2\pi^2}\ \xi_c\xi_t\ m_c^2\
\log\bfrac{m_t^2}{m_c^2},\cr }} substituting $c_1=-4G_F/\sqrt2$.

\begfig 0 cm
\centerline{\hbox{
\epsfxsize=3.5cm\epsffile{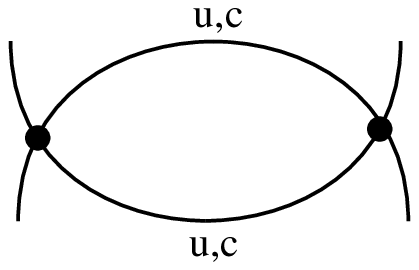}}}
\figure{\figscale}{The infinite part of this graph contributes to the
renormalization group scaling of the $\Delta S=2$ amplitude}
\endfig

\titleb{}{Matching at $m_c$}
Finally, one integrates out the $c$ quark. This is virtually identical to the
matching condition at the $t$-quark scale, and gives
\eqn\matchc{
c_2(\mu=m_c-0)= c_2(\mu=m_c+0) +{G_F^2\over 4\pi^2} \left[ \xi_c^2 m_c^2 + 2
\xi_c\xi_u m_c^2\right].  } Combining \matcht--\matchc\ reproduces the
the
box diagram computation \inamilimit.

There are some important features of the $\Delta S=2$ computation which are
generic to any effective field theory computation. (i) The contributions
proportional to the heaviest mass scale $m_t$ arise from matching conditions
at
that scale. (ii) contributions proportional to lower mass scales (such as
$m_c$) arise from matching at the scale $m_c$, and also from from matching
at scales larger than $m_c$ (such as $m_t$).
(iii)
Contributions proportional to logarithms of two scales arise from
renormalization group evolution between the two scales.

It seems that the effective field theory method is much more complicated than
directly computing the original box diagram in
Fig.~\figbox. The effective theory method has broken the computation of the box
diagram into several steps.  The computations involved at each step in the
effective field theory are much simpler than the box diagram calculation. The
box diagram involves several different mass scales in the internal
propagators,
which leads to complicated Feynman parameter integrals that must be
evaluated. The matching condition computations in the effective field theory
each involve only a single mass scale, and are much simpler. One can contrast
the full answer \inamilimit\ with the individual pieces of the effective
field theory calculation in \matcht--\matchc. Furthermore, in the
effective field theory calculation it is trivial to include the leading
logarithmic QCD corrections to the $\Delta S =2$ amplitude. The corresponding
computation in the full theory is far more difficult, and involves computing
two loop diagrams such as the one in Fig.~\fig\figboxcorr{figboxcorr}.

\begfig 0 cm
\centerline{\hbox{
\epsfxsize=3.5cm\epsffile{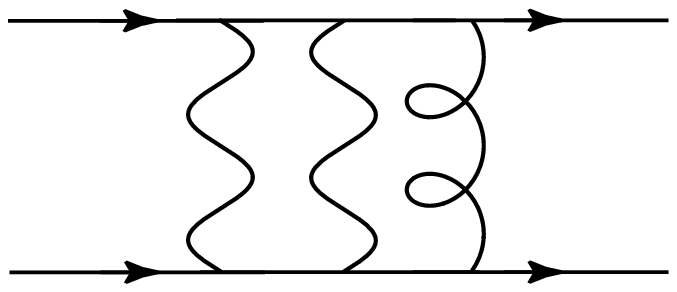}}}
\figure{\figboxcorr}{A QCD radiative correction to the box diagram}
\endfig

\titleb{}{QCD Corrections}
The leading logarithmic corrections to the $\Delta S=2$ amplitude sum all
corrections of the form $(\alpha_s \log r)^n$, where $r$ is large ratio of
scales such as $M_W/m_c$, but neglect corrections of the form
$\alpha_s(\alpha_s \log r)^n$. The QCD corrections to the matching condition
only involve a single scale, and do not have any large logarithms. For
example,
the matching condition at the scale $m_t$ only involves corrections that
depend
on $\alpha_s(\mu)$ and $\log m_t/\mu$. Evaluating these corrections by setting
the $\bar{MS}$ parameter $\mu=m_t$ implies that there is no leading
logarithmic
correction to the matching condition. The only leading logarithmic QCD
corrections arise from renormalization group scaling between different
scales. This computation is straightforward, and only involves the infinite
parts of one loop diagrams. The renormalization group equation \rgct\ is
replaced by
\eqn\newrg{
\mu {d\over d\mu}\, c_2(\mu) = \bfrac{1}{8\pi^2}\ m_c^2(\mu)\, c_1^2(\mu)\
\xi_c\xi_t +
\gamma_2(\mu)\ c_2(\mu),
} where $m_c$ has been replaced by the running mass, $c_1$ has been replaced
by
the running coupling $c_1(\mu)$, and $\gamma_2$ is the anomalous dimension
\eqn\gamii{
\gamma_2 = \bfrac{\alpha_s(\mu)}\pi,
} of the $\Delta S=2$ operator $(\bar d\, \gamma^\mu\,P_L\,s)\, (\bar d\,
\gamma_\mu\,P_L\,s)$, which can be obtained from the infinite part of
Fig.~\fig\figdsii{figdsii}. The running mass $m_c(\mu)$ satisfies the
renormalization group equation
\eqn\gamm{
\mu {d\over d\mu}\, m_c(\mu) =\gamma_m\ m_c(\mu) =
 -\bfrac{2\alpha_s(\mu)}\pi\ m_c(\mu).  } If $c_1$ satisfies a simple
renormalization group equation of the form
\eqn\gamisimp{
\mu {d\over d\mu}\, c_1(\mu) = \gamma_1\ c_1(\mu),
}
one can solve \newrg--\gamisimp\ to obtain the QCD corrected value for
$c_2(\mu)$. At one loop, it is convenient to define $b$ and $\hat\gamma_i$
\eqn\bdef{
\mu {d\over d\mu}\, g = \beta(g) = - b \bfrac{g^3}{16\pi^2} + \ldots,
} and
\eqn\XVvi{
\gamma_i = \hat \gamma_i\ \bfrac{g^2}{16\pi^2} + \ldots,
}
for $i=1,2,m$. One can then solve \gamm\ and \gamisimp,
\eqn\misolve{\eqalign{
m_c(\mu) &= m_c(\mu^\prime) \left[\bfrac{g(\mu)}{g(\mu^\prime)}\right]^{-
\hat\gamma_m/b}=\left[\bfrac{\alpha_s(\mu^\prime)}{\alpha_s(\mu)}\right]^{
\hat\gamma_m/2 b},\cr
c_1(\mu) &= c_1(\mu^\prime)
\left[\bfrac{g(\mu)}{g(\mu^\prime)}\right]^{-\hat\gamma_1/b}
=\left[\bfrac{\alpha_s(\mu^\prime)}{\alpha_s(\mu)}\right]^{
\hat\gamma_1/2 b}.
}} Substituting \misolve\ into \newrg\ and integrating gives
\eqn\wrong{\eqalign{
c_2(m_c)&=c_2(m_t) \left[\bfrac{\alpha_s(m_t)} {\alpha_s(m_c)}
\right]^{\hat\gamma_2/2 b}\cr
&+ \bfrac{m_c^2(m_t)\ c_1^2(m_t)} {g(m_t)^2
\left(2+2\hat\gamma_1/b+2\hat\gamma_m/b-\hat\gamma_2/b\right)}
\left[\left(\bfrac{\alpha_s(m_t)} {\alpha_s(m_c)}\right)^
{2+2\hat\gamma_1/b+2\hat\gamma_m/b}\right. \cr
&\qquad\qquad\qquad\qquad - \left.
\left(\bfrac{\alpha_s(m_t)} {\alpha_s(m_c)}\right)^{\hat\gamma_2/2 b}
\right].
}}

\begfig 0 cm
\centerline{\hbox{
\epsfxsize=3cm\epsffile{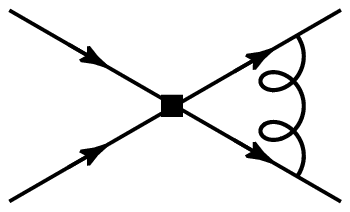}}}
\figure{\figdsii}{Graph contributing to the anomalous dimension of the
$\Delta s=2$ operator $(\bar d\, \gamma^\mu \, P_L\,s)\, (\bar d\,
\gamma_\mu \, P_L\,s)$}
\endfig

The actual computation of these effects in the standard model is more involved,
because the $\Delta S=1$ Lagrangian does not satisfy a simple renormalization
group equation of the form \gamisimp. There is operator mixing, and
\gamisimp\ is replaced by a matrix equation.  Nevertheless, it is possible
to compute the results using an effective field theory method, though the
final
form of the answer is more complicated than \wrong. The reader is referred
to the papers by Gilman and Wise for details.  The computation of QCD
corrections in the full theory is far more complicated, and has never been
done.

To compare the advantages and disadvantages of the full and effective theory
computation, let us concentrate only on the $m_t$ part of the $\Delta S=2$
amplitude.  The effective field theory computation gives the $\Delta S=2$
amplitude as an expansion in powers of $m_t/M_W$, and we have computed the
leading term in \inamilimit. The general form of the effective field
theory
result is
\eqn\ansexp{
{\rm answer} = \left(\bfrac{m_t}{M_W}\right)^2\left(\bfrac{\alpha_s(M_W)}
{\alpha_s(m_t)}\right)^{\gamma_2/2b} +
\left(\bfrac{m_t}{M_W}\right)^4\left(\bfrac{\alpha_s(M_W)}
{\alpha_s(m_t)}\right)^{\gamma_4/2b} + \ldots } where $\gamma_i$ are the
anomalous dimensions of the dimension six, eight, etc. operators. (For
example,
compare with \wrong.) Evaluating each of these anomalous dimension is a
separate computation. Equation~\ansexp\ is useful if there is a large ratio of
scales, $m_t/M_W \ll 1$, so that one only needs a few terms in the expansion
\ansexp. The full theory computation \inamilim\ sums up the entire series,
and gives an answer of the form
\eqn\ansfull{
{\rm answer} = f(m_t/M_W), } which is valid for any value of the ratio
$m_t/M_W$. The computations involved in \ansfull\ are necessarily more
complicated than those for the effective field theory, because one obtains the
entire functional form of the answer, rather than the first few terms in a
series expansion. However, it is not possible to compute the leading
logarithmic QCD corrections to \ansfull, since each term in the expansion
has a different anomalous dimension. For the $c$ quark, it is more important
to
sum the leading QCD corrections, than to include higher order terms in
$m_c/M_W\sim1/50$, and the effective theory method is useful. The recently
measured value of the top quark mass indicates that the ratio $m_t/M_W\sim
2$. In this case, it is more important to retain the entire form of the
$m_t/M_W$ dependence, than to include the QCD radiative corrections.  The way
the calculation is done in practice is to integrate out the $t$-quark and
$W$-boson together at some scale $\mu$ which is comparable to both $m_t$ and
$M_W$, and then use an effective theory to scale down to $m_c$ so as to
include
the QCD corrections between $\{ M_W,m_t \}$ and $m_c$. Clearly, the ideal
procedure would be to retain the entire functional form \ansfull, as well
as the entire QCD radiative correction. This has been done in a toy model
using
a non-local effective Lagrangian, but it is not known how to do this in
general.

A very different example where an infinite set of anomalous dimensions can be
computed is the QCD evolution of parton structure functions.  In QCD, the
Altarelli-Parisi splitting functions for the parton distribution functions
contain the same information as the infinite set of anomalous dimensions of
the
twist-two operators. The distribution functions can be written as matrix
elements of non-local operators, and the one-loop anomalous dimension is a
function, whose moments give the anomalous dimensions of the infinite tower of
twist two operators.

\titlea{9}{The Non-linear Sigma Model}
The previous results discussed effective field theories in the perturbative
regime, where one could compute the effective Lagrangian from the full theory
in a systematic perturbative expansion. One can also apply effective field
theory ideas to situations where one can not derive the effective Lagrangian
from the full theory directly. The classic example of this is the use of
non-linear sigma models to study spontaneously broken global symmetries, and
in particular, the use of chiral Lagrangians to study pion interactions in
QCD.

Consider first the linear sigma model with Lagrangian
\eqn\linl{
\CL = \frac12 \partial_\mu \bfm \phi\,\cdot \partial^\mu \bfm\phi - \lambda
(\bfm\phi\cdot\bfm\phi-v^2)^2, } where $\bfm\phi=(\phi_1,\ldots,\phi_N)$ is a
real $N$-component scalar field. This theory will illustrate some ideas
which will be needed for the study of chiral symmetry breaking in QCD.
The Lagrangian \linl\ has a global $O(N)$ symmetry under which $\bfm\phi$
transforms as an $O(N)$ vector. The potential has been chosen so that it is
minimized for $\abs{\bfm\phi}=v$. The set of field configurations where
$\abs{\bfm\phi}=v$ is known as the vacuum manifold, and in our example, it is
the set of points $\bfm\phi=(\phi_1,\ldots,\phi_N)$, with $\phi_1^2+\phi_2^2
+\ldots +\phi_N^2=v^2$, i.e. it is the $N-1$ dimensional sphere $S^{N-1}$. The
$O(N)$ symmetry can be used to rotate the vector $\vev{\bfm\phi}$ to a
standard
direction, which can be chosen to be $(0,0,\ldots,v)$, the north pole of the
sphere. The vacuum of the Lagrangian has spontaneously broken the $O(N)$
symmetry down to the $O(N-1)$ subgroup which acts on the first $N-1$
components. The other generators of $O(N)$ do not leave $(0,0,\ldots,v)$
invariant. $O(N)$ has $N(N-1)/2$ generators, so the number of Goldstone bosons
is equal to the number of broken generators, $N(N-1)/2 - (N-1)(N-2)/2 =
N-1$. The $N-1$ Goldstone bosons correspond to rotations of the vector
$\bfm\phi$, which leave its length unchanged. The potential energy $V$ is
unchanged under rotations of $\bfm\phi$, so these modes are massless.  The
remaining mode is a radial excitation which changes the length of $\bfm\phi$,
and produces a massive excitation, with mass $m_H=
\sqrt{8\lambda}\, v$.

It is convenient to switch to ``polar coordinates'', and define
\eqn\polarco{
\bfm\phi=\left(\rho+v\right)\ e^{i \sum_s X^s\cdot \pi^s}
\pmatrix{0\cr0\cr.\cr.\cr.\cr1\cr},
} where $X^s, s=1,\ldots,N-1$ are $N-1$ broken generators, and $\pi^s$ and
$\rho$ are a new basis for the $N$ fields. This change of variables is only
well-defined for small angles $\pi^s$.  The Lagrangian in terms of the new
fields is
\eqn\polarl{
\CL = \bfrac12 \partial_\mu \rho \partial^\mu \rho - \lambda\left(\rho^2 +
2\rho v\right)^2 +\bfrac12 \left(\rho+v\right)^2 \left[\partial_\mu e^{-i
\sum_s X^s\cdot \pi^s}\partial^\mu e^{i \sum_s X^s\cdot \pi^s}
\right]_{NN},
} where $[\ ]_{NN}$ is the $NN$ element of the matrix.  At energies small
compared to the radial excitation mass $\sqrt{8\lambda}\, v$, the $\rho$ field
can be neglected, and the Lagrangian reduces to
\eqn\polarlapprox{
\CL = \bfrac12 v^2 \left[\partial_\mu
e^{-i \sum_s X^s\cdot \pi^s}\partial^\mu e^{i \sum_s X^s\cdot \pi^s}
\right]_{NN},
} which describes the self-interactions of the Goldstone bosons.

There are some generic features of Goldstone boson interactions that are easy
to understand:
\medskip
\item{1.} The Goldstone boson fields are derivatively coupled.
The Goldstone bosons describe the local orientation of the $\bfm\phi$ field. A
constant Goldstone boson field is a $\bfm\phi$ field that has been rotated by
the same angle everywhere in spacetime, and corresponds to a vacuum that is
equivalent to the standard vacuum $\vev{\bfm\phi}=(0,0,\ldots,1)$. Thus the
Lagrangian must be independent of $\pi^s$ when $\pi^s$ is a constant, so only
gradients of $\pi^s$ appear in the Lagrangian.
\item{2.} The effective Lagrangian describes a theory of {\it weakly}
interacting Goldstone bosons at low energy. The Goldstone boson couplings are
proportional to their momentum, and so vanish for low-momentum Goldstone
bosons.
\item{3.} The Goldstone boson Lagrangian is non-linear in the Goldstone
boson fields. The Goldstone boson Lagrangian describes the dynamics of fields
constrained to live on the vacuum manifold. The constraint equation,
$\phi_1^2+\phi_2^2 +\ldots +\phi_N^2=v^2$, is non-linear, and leads to a
non-linear Lagrangian.
\item{4.} The vacuum manifold is generically curved (like our
sphere $S^{N-1}$), and does not have a set of global coordinates. The $\pi^s$
coordinates defined in \polarco\ only make sense for small fluctuations of
the Goldstone boson fields about the north pole, which is adequate for
perturbation theory. For studying non-perturbative effects or global
properties, it is better not to introduce the angular coordinates, but to
write
the Lagrangian directly in terms of fields that take values on the vacuum
manifold, $\pi(x)\in S^{N-1}$.
\item{5.} The amplitude for the broken symmetry currents to produce a
Goldstone boson from the vacuum is proportional to the symmetry breaking
strength $v$.

\titlea{10}{The CCWZ Formalism}
The general formalism for effective Lagrangians for spontaneously broken
symmetries was worked out by Callan, Coleman, Wess, and Zumino. Consider a
theory in which a global symmetry group $G$ is spontaneously broken to a
subgroup $H$. The vacuum manifold is the coset space $G/H$. In our example,
$G=O(N)$, $H=O(N-1)$, and $G/H=O(N)/O(N-1) = S^{N-1}$.

We would like to choose a set of coordinates which describe the local
orientation of the vacuum for small fluctuations about the standard vacuum
configuration. Let $\Xi(x)\in G$ be the rotation matrix that transforms the
standard vacuum configuration to the local field configuration. The matrix
$\Xi$ is not unique: $\Xi h$, where $h\in H$, gives the same field
configuration, since the standard vacuum is invariant under $H$
transformations. In our example, one can describe the direction of the vector
$\bfm\phi$ by giving the $O(N)$ matrix $\Xi$, where
\eqn\XXi{
\bfm \phi(x) = \Xi(x) \pmatrix{0\cr0\cr.\cr.\cr.\cr v\cr}.
}
The same configuration $\bfm\phi(x)$ can also be described by $\Xi(x) h(x)$,
where $h(x)$ is a matrix of the form
\eqn\Xii{
h(x) = \pmatrix{h'(x) & 0 \cr 0&1\cr},
}
with $h'(x)$ an arbitrary $O(N-1)$ matrix, since
\eqn\Xiii{
\pmatrix{h'(x) & 0 \cr 0&1\cr} \pmatrix{0\cr0\cr.\cr.\cr.\cr v\cr}
= \pmatrix{0\cr0\cr.\cr.\cr.\cr v\cr}.
}
The CCWZ prescription is to pick a set of broken generators $X$, and choose
\eqn\ccwzform{
\Xi(x) = e^{i X\cdot\pi(x)}.
} Consider the $O(N)$ theory for $N=3$, which is the theory of a vector
$\bfm\phi$ in three-dimensions, and so is easy to visualize.  The symmetry
group $G$ is the group $G=O(3)$ of rotations in three-space. The standard
vacuum configuration $\vev{\bfm \phi}$ can be chosen to be $\bfm\phi$ pointing
towards the north pole $N$, and the unbroken symmetry group $H=O(2)=U(1)$ is
rotations about the axis $ON$, where $O$ is the center of the sphere (see
Fig.~\fig\sigmafig{The vacuum manifold for the $O(3)$ sigma model. The standard
configuration $\bfm \phi$ is along $ON$. Under the transformation $g$, $A$
gets mapped to $B$.}). The group
generators are $J_1, J_2,J_3$, and the unbroken generator is $J_3$, where
$J_k$
generate rotations about the $k$th axis.  The CCWZ prescription is to choose
\eqn\xichoice{
\Xi(x) = e^{i \left[J_1 \pi(x) + J_2 \pi_2(x)\right]}
}
to represent $\bfm \phi$ along $OA$. The matrix $\Xi$ rotates a vector
pointing along the $ON$ axis to $\bfm\phi=OA$ by
rotating along a line of longitude.

\begfig 0 cm
\centerline{\hbox{
\epsfxsize=4cm\epsffile{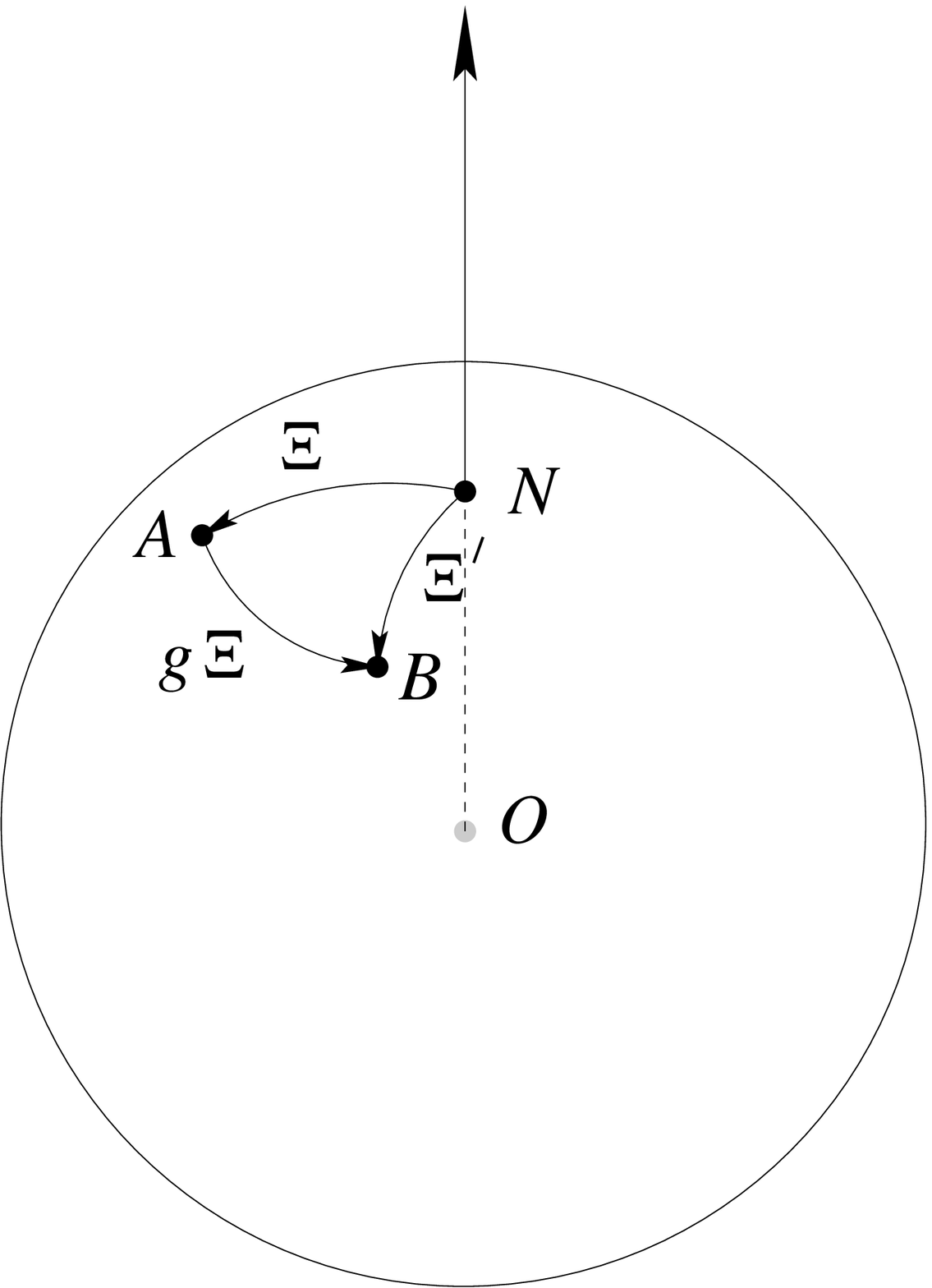}}}
\figure{\sigmafig}{The vacuum manifold for the $O(3)$ sigma model. The standard
configuration $\bfm \phi$ is along $ON$. Under the transformation $g$, $A$
gets mapped to $B$}
\endfig

Under a global symmetry transformation $g$, the matrix $\Xi(x)$ is transformed
to the new matrix $g\Xi(x)$, since $\bfm\phi(x)\rightarrow g\bfm\phi(x)$.
(Note that $g$ is a global transformation, and does not depend on
$x$.) The new matrix $g\Xi(x)$ is no longer in standard form, \ccwzform,
but can be written as
\eqn\curvedh{
g\ \Xi = \Xi'\ h, } since two matrices $g\,\Xi$ and $\Xi'$ which describe the
same field configuration differ by an $H$ transformation. That $h$ is
non-trivial is a well-known property of rotations in three dimensions. Take an
object and rotate it from $N$ to $A$ and then to $B$. This transformation is
not the same as a direct rotation from $N$ to $B$, but can be written as a
rotation about $ON$, followed by a rotation from $N$ to $B$.  The
transformation $h$ in \curvedh\ is non-trivial because the Goldstone boson
manifold $G/H$ is curved.

The transformation \curvedh\ is usually written as
\eqn\ccwztrans{
\Xi(x) \rightarrow g\ \Xi(x)\ h^{-1}(g,\Xi(x)),
} where we have made clear the implicit dependence of $h$ on $x$ through its
dependence on $g$ and $\Xi(x)$. Equations~\ccwzform\ and \ccwztrans\ give the
CCWZ
choice for the Goldstone boson field, and its transformation law. Any other
choice gives the same results for all observables, such as the $S$-matrix, but
does not give the same off-shell Green functions.

\titlea{11}{The QCD Chiral Lagrangian}
The CCWZ formalism can now be applied to QCD. In the limit that the $u$, $d$
and $s$ quark masses are neglected, the QCD Lagrangian has a
$SU(3)_L\times SU(3)_R$ chiral symmetry under which the left- and right-handed
quark fields transform independently,
\eqn\XIi{
\psi_L(x) \rightarrow L\ \psi_L(x),\qquad
\psi_R(x) \rightarrow R\ \psi_R(x),
}
where
\eqn\XIii{
\psi = \pmatrix{u\cr d \cr s\cr}.
}
The $SU(3)_L\times SU(3)_R$ chiral symmetry is spontaneously broken to the
vector $SU(3)$ subgroup by the $\vev{\bar\psi\psi}$ condensate. The symmetry
group is $G=SU(3)_L\times SU(3)_R$, the unbroken group is $H=SU(3)_V$, and the
Goldstone boson manifold is the coset space $SU(3)_L\times SU(3)_R/SU(3)_V$
which is isomorphic to $SU(3)$. The generators of $G$ are $T^a_L$ and $T^a_R$
which act on left and right handed quarks respectively, and the generators of
$H$ are the flavor generators $T^a=T^a_L+T^a_R$. There are two commonly used
bases for the QCD chiral Lagrangian, the $\xi$-basis and the $\Sigma$-basis,
and we will consider them both. There are many simplifications that occur for
QCD because the coset space $G/H$ is isomorphic to a Lie group. This is not
true in general; in the $O(N)$ model, the space $S^{N-1}$ is not isomorphic to
a Lie group for $N\not=4$.

\titleb{}{The $\xi$-basis}
The unbroken generators of $H$ plus the broken generators $X$ span the space
of
all symmetry generators of $G$. One choice of broken generators is to pick
$X^a=T^a_L-T^a_R$. Let the $SU(3)_L\times SU(3)_R$ transformation be
represented in block diagonal form,
\eqn\gblock{
g=\left[\matrix{L&0\cr 0&R\cr}\right], } where $L$ and $R$ are the $SU(3)_L$
and $SU(3)_R$ transformations, respectively. The unbroken transformation have
the form \gblock\ with $L=R=U$,
\eqn\hblock{
h = \left[\matrix{U&0\cr 0&U\cr}\right].  } The $\Xi$ field is then defined
using the CCWZ prescription \ccwzform
\eqn\XIv{
\Xi(x) = e^{i X\cdot \pi(x)} = \exp i \left[\matrix{T\cdot\pi&0\cr 0& -
T\cdot\pi\cr}\right] = \left[\matrix{\xi(x)&0\cr 0&\xi^\dagger(x)\cr}
\right],
}
where
\eqn\XIvi{
\xi = e^{i T\cdot \pi}
}
denotes the upper block of $\Xi(x)$.  The transformation rule \ccwztrans\
gives
\eqn\XIvii{
\left[\matrix{\xi(x)&0\cr 0&\xi^\dagger(x)\cr}\right]
\rightarrow
\left[\matrix{L&0\cr 0&R\cr}\right]
\left[\matrix{\xi(x)&0\cr 0&\xi^\dagger(x)\cr}\right]
\left[\matrix{U^{-1}&0\cr 0&U^{-1}\cr}\right].
}
This gives the transformation law for $\xi$,
\eqn\xitrans{
\xi(x) \rightarrow L\ \xi(x)\ U^{-1}(x) = U(x)\ \xi(x)\ R^\dagger,
} which defines $U$ in terms of $L$, $R$, and $\xi$.

\titleb{}{The $\Sigma$ basis}
The $\Sigma$-basis is obtained from the CCWZ prescription using $X^a=T^a_L$
for
the broken generators. In this case, \ccwzform\ gives
\eqn\XIix{
\Xi(x) = e^{i X\cdot \pi(x)} = \exp i \left[\matrix{T\cdot\pi&0\cr 0&
0\cr}\right] = \left[\matrix{\Sigma(x)&0\cr0&1\cr}\right]
}
where
\eqn\XIx{
\Sigma = e^{i T \cdot \pi}
}
denotes the upper block of $\Xi(x)$.  The transformation law \ccwztrans\
is
\eqn\XIxi{
\left[\matrix{\Sigma(x)&0\cr 0&1\cr}\right]
\rightarrow
\left[\matrix{L&0\cr 0&R\cr}\right]
\left[\matrix{\Sigma(x)&0\cr 0&1\cr}\right]
\left[\matrix{U^{-1}&0\cr 0&U^{-1}\cr}\right],
}
which gives $U=R$, and
\eqn\sigtrans{
\Sigma(x) \rightarrow L\ \Sigma(x)\ R^\dagger.
} Comparing with \xitrans, one sees that $\Sigma$ and $\xi$ are related by
\eqn\sigxi{
\Sigma(x)=\xi^2(x).
}

\titleb{}{The Lagrangian}
The Goldstone boson fields are angular variables, and are dimensionless. When
writing down effective Lagrangians in field theory, it is convenient to use
fields which have mass dimension one, as for any other spin-zero boson
field. The standard choice is to use
\eqn\XIxii{
\xi = e^{iT\cdot\pi/f},\qquad \Sigma=e^{2iT\cdot \pi/f},
}
where $f\sim93$~MeV is the pion decay constant. The $\pi$ matrix is
\eqn\XIxiv{
\bfm\pi=\pi^a T^a,
}
where the group generators have the usual normalization $\tr T^a T^b =
\delta^{ab}/2$,
\eqn\pifield{
\bfm\pi = {1\over\sqrt2}\left[\matrix{\bfrac1{\sqrt2}\pi^0 +
\bfrac1{\sqrt6}\eta&\pi^+&K^+\cr\pi^-&-\bfrac1{\sqrt2}\pi^0 +\bfrac1{\sqrt6}
\eta&K^0\cr K^-&\bar K^0&-\bfrac2{\sqrt6}\eta}\right].
}

The low energy effective Lagrangian for QCD is the most general possible
Lagrangian consistent with spontaneously broken $SU(3)\times SU(3)$
symmetry. Unlike our weak interaction example, one cannot simply compute the
effective Lagrangian directly from the original QCD Lagrangian. The connection
between the original and effective theories is non-perturbative. The effective
Lagrangian has an infinite set of unknown parameters, but we will see that it
can still be used to obtain non-trivial predictions for experimentally
measured
quantities.

It is easy to construct the most general Lagrangian invariant under the
transformation $\Sigma \rightarrow L \Sigma R^\dagger$. The most general
invariant term with no derivatives must be the product of terms of the form
$\Tr \Sigma \Sigma^\dagger\ldots
\Sigma \Sigma^\dagger$, where $\Sigma$ and $\Sigma^\dagger$'s
alternate. However, $\Sigma \Sigma^\dagger=1$, so all such terms are constant,
and independent of the pion fields. This is just our old result that all
Goldstone bosons are derivatively coupled. The only invariant term with two
derivatives is
\eqn\lii{
\CL_2 = \bfrac{f^2}4 \Tr\, \partial_\mu \Sigma\partial^\mu \Sigma^\dagger.
} Expanding $\Sigma$ in a power series in the pion field gives
\eqn\liiexp{
\CL_2 = \Tr\, \partial_\mu \pi \partial^\mu \pi + \bfrac 1 {3f^2}
\Tr\,\left[\pi,\partial_\mu\pi\right]^2 + \ldots\ \ .
} The coefficient of the two-derivative term in \lii\ is fixed by
requiring
that the kinetic term for the pions in \liiexp\ has the standard
normalization for scalar fields. The Lagrangian \lii\ only has terms with
an even number of pions, since the pion is a pseudoscalar.  The Lagrangian
\lii\ determines all the multi-pion scattering amplitudes to order $p^2$
in
terms of a single constant $f$. For example, the $\pi-\pi$ scattering
amplitude
is given by the term $\Tr\left[\pi,\partial_\mu\pi\right]^2/3f^2$, etc.

\titleb{}{The Chiral Currents}
Noether's theorem can be used to compute the $SU(3)_L$ and $SU(3)_R$ currents.
If a Lagrangian $\CL$ is invariant under an infinitesimal global symmetry
transformation with parameter $\epsilon$, the current $j_\mu$ is given by
computing the change of the Lagrangian when one makes the same transformation,
with $\epsilon$ a function of $x$,
\eqn\XIxvii{
\delta \CL = \partial_\mu\epsilon(x)\ j^\mu(x).
}
The infinitesimal form of the $SU(3)_L$ transformation $\Sigma\rightarrow L
\Sigma$ is
\eqn\ltrans{
\Sigma \rightarrow \Sigma + i \epsilon_L^a\, T^a\Sigma,
} where $L=\exp i \epsilon_L^a T^a \approx 1 + i \epsilon_L^a T^a + \ldots$.
The change in \lii\ under \ltrans\ is
\eqn\XIxviii{
\delta\CL =  \partial_\mu \epsilon_L^a\ \Tr\, T^a \Sigma \partial^\mu
\Sigma^\dagger
}
so that the $SU(3)_L$ currents are
\eqn\XIxix{
j^\mu_L = \bfrac i 2\, f^2 \,\Tr\, T^a \Sigma \partial^\mu \Sigma^\dagger.
}
The right handed currents are obtained by applying the parity transformation,
$\pi(x)\rightarrow -\pi(-x)$ or by making an infinitesimal $SU(3)_R$
transformation, so that
\eqn\XIxx{
j^\mu_R = \bfrac i 2\, f^2\,\Tr\, T^a\Sigma^\dagger \partial^\mu\Sigma.
}
The axial current has the expansion
\eqn\jaxial{
j^{\mu a}_A = j^{\mu a}_R - j^{\mu a}_L = -f \partial^\mu \pi^a + \ldots } The
matrix element $\bra 0 j^{\mu a}_A \ket {\pi^b} = i f p^\mu\delta^{ab}$, so
that $f$ is the pion decay constant. The experimental value of the $\pi$ decay
rate, $\pi\rightarrow \mu \bar\nu$ determines $f\approx 93$~MeV.

The low-energy effective theory of the weak interactions is an expansion in
some low mass scale (such as $m_c$ or $\lqcd$) over $M_W$. The QCD chiral
Lagrangian is an expansion in derivatives, and so is an expansion in
$p/\lchi$.
The pion couplings are weak, as long as the pion momentum is small compared
with $\lchi$. There are two important questions that have to be answered
before
one can use the effective Lagrangian: (i) What terms in the effective
Lagrangian are required to compute to a given order in $p/\lchi$? (ii) What is
the value of $\lchi$? Then one has an estimate of the neglected higher-order
terms in the expansion, and the energy at which the effective theory breaks
down.

It is useful to eliminate all redundant terms in the
effective Lagrangian. One can often eliminate many terms in the effective
Lagrangian by making suitable field redefinitions.  Field redefinitions are
not
very useful in renormalizable field theories, because they make renormalizable
Lagrangians look superficially non-renormalizable. For example, a field
redefinition
\eqn\fredef{
\phi \rightarrow \phi + \epsilon \phi^2,
} turns
\eqn\renlag{
\CL = \bfrac12 \partial_\mu\phi\,\partial^\mu\phi - \bfrac12 m^2\phi^2
-\lambda \phi^4 } into
\eqn\rennew{
\CL= \bfrac12 \partial_\mu\phi\, \partial^\mu\phi - \bfrac12 m^2\phi^2
-\lambda \phi^4 + \epsilon\left(2\phi\, \partial_\mu\phi \, \partial^\mu\phi -
m^2\phi^3 - 4 \lambda \phi^5\right) + \CO\left(\epsilon^2\right), } which
looks
superficially like a non-renormalizable interaction.  Equations~\rennew\ and
\renlag\
define identical theories, and the field redefinition \fredef\ has turned
a
simple Lagrangian into a more complicated one. However, in the case of
non-renormalizable theories which contain an infinite number of terms, one can
use field redefinitions to eliminate many higher-order terms in the effective
Lagrangian (see Ref.~7). The way this is usually done in practice is to use
the
equations of motion derived from the lowest-order terms in the effective
Lagrangian to simplify or eliminate higher order terms.

\titleb{}{Weinberg's Power Counting Argument}
The QCD chiral Lagrangian is
\eqn\XIxxx{
\CL=\sum_k \CL_k,
}
where $\CL_2$, $\CL_4$, etc.\ are the terms in the Lagrangian with two
derivatives, four derivatives, and so on. Consider an arbitrary loop graph,
such
as the one in
Fig.~\fig\pprandom{Random graph}. It
contains $m_2$ interaction vertices that come from terms in $\CL_2$, $m_4$
interaction vertices from terms in $\CL_4$, etc. The general form of the
diagram is
\eqn\genform{
\CA = \int \left(d^4 p\right)^L \bfrac 1 {\left(p^2\right)^I}
\prod_k \left( p^k \right)^{m_k},
}
where $L$ is the number of loops, $I$ is the number of internal lines, and $p$
represents a generic momentum. The factors are easy to understand: there is a
$d^4 p$ integral for each loop, each internal boson propagator is $1/p^2$, and
each vertex in $\CL_k$ gives a factor of $p^k$. In a mass-independent
subtraction scheme, the only dimensional parameters are the momenta $p$. Thus
the amplitude $\CA$ must have the form $\CA \sim p^D$, where
\eqn\done{
D = 4 L - 2I + \sum_k\ k\ m_k, } from \genform. For any Feynman graph, one
can show that
\eqn\euler{
V-I+L=1, } where $V$ is the number of vertices, $I$ is the number of internal
lines, and $L$ is the number of loops. Combining \done,\euler, and using
$V=\sum_k m_k$, one gets
\eqn\weini{
D = 2 + 2L + \sum_k\left(k-2\right) m_k.  } The chiral Lagrangian starts at
order $p^2$, so $k\ge2$, and all terms in \weini\ are non-negative. As a
result, only a finite number of terms in the effective Lagrangian are needed
to
work to a fixed order in $p$, and the chiral Lagrangian acts like a
renormalizable field theory. For example, to compute the scattering amplitudes
to order $p^4$, one needs
\eqn\XIIi{
4 = 2 + 2L + \sum_k\left(k-2\right) m_k,
}
which has the solutions $L=0$, $m_4=1$, $m_{k>4}=0$, or $L=1$ and
$m_{k>2}=0$. That is, one only needs to consider tree level diagrams with one
insertion of $\CL_4$, or one-loop graphs with the lowest order Lagrangian
$\CL_2$ to compute all scattering amplitudes to order $p^4$.

\begfig 0 cm
\centerline{\hbox{
\epsffile{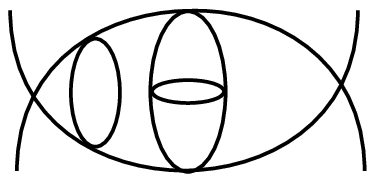}}}
\figure{\pprandom}{A loop graph for multipion interactions}
\endfig

\titleb{}{Naive Dimensional Analysis}
Consider the $\pi-\pi$ scattering amplitude to order $p^4$. The power counting
argument implies that there are two contributions to this: a tree level graph
with one insertion of $\CL_4$, and a loop graph using $\CL_2$. The loop graph
is of order
\eqn\loopest{
I\sim\int {d^4k\over\left(2\pi\right)^4}\ \bfrac{k^2}{f^2}\ \bfrac{k^2}{f^2}\
\bfrac{1}{k^4},
} where $1/k^4$ is from the two internal propagators, and each four-pion
interaction vertex is of order $k^2/f^2$, from \liiexp. Here $k$ denotes a
generic internal momentum in the Feynman diagram.
Estimating this
integral gives
\eqn\iest{
I\sim \bfrac{p^4}{16\pi^2}\,\bfrac1{f^4}\ \log\mu, } where $\mu$ is the
$\bar{MS}$ renormalization scale, and $p$ represents a generic external
momentum. A four derivative operator in the Lagrangian
of the form
\eqn\fourder{
a \tr\, \partial_\mu \Sigma\partial^\mu \Sigma^\dagger\partial_\nu \Sigma
\partial^\nu \Sigma^\dagger,
} produces a four-pion interaction of order $a p^4/f^4$ when one expands the
$\Sigma$ field in a power series in $\pi/f$. The total four-pion amplitude,
which is the sum of the tree and loop graphs, is $\mu$-independent.  A shift
in
the renormalization scale $\mu$ is compensated for by a corresponding shift in
$a$. A change in $\mu$ of order one produces a shift in $a$ of order $ \delta
a\sim {1/ 16\pi^2}$.  Generically, $a$ must be at least as big as $\delta a$,
\eqn\ashift{
\abs{a} \ga \abs{\delta a} \sim {1\over 16\pi^2}, } because a shift in the
renormalization point of order one produces a shift in $a$ of this size. Write
the effective Lagrangian as
\eqn\lstd{
\CL = \bfrac {f^2} 4\left[ \tr \partial_\mu \Sigma\partial^\mu
\Sigma^\dagger + \bfrac1{\lchi^2} \CL_4 + \bfrac{1}{\lchi^4}\CL_6 + \ldots
\right]
} where $1/\lchi$ is the expansion parameter of the effective Lagrangian,
i.e. \lstd\ gives an expansion for scattering amplitudes in powers of
$p/\lchi$. The estimate \ashift\ for the size of the four derivative term
implies that
\eqn\lchiest{
\lchi \la 4\pi f.
} One can show that a similar estimate holds for all the higher order terms in
$\CL$, i.e. the six derivative term has a coefficient of order $1/\lchi^4$,
etc. Numerous calculations suggest that in QCD, the inequality \lchiest\
can be replaced by the estimate
\eqn\lvalue{
\lchi \sim 4\pi f \sim 1\ {\rm GeV},
} for the expansion parameter of the effective Lagrangian. This parameter is
large enough that one can apply chiral Lagrangians to low energy processes
involving pions and kaons. If the expansion parameter were $f$ instead of
$4\pi
f$, chiral Lagrangians would not be useful even for pions, since $m_\pi > f$.

The naive dimensional analysis estimate equivalent to \lstd\ is that a
term
in the Lagrangian has the form
\eqn\ndest{
f^2 \lchi^2 \left(\bfrac{\pi}f\right)^n \left(\bfrac{\partial}{
\lchi}\right)^m,
} as can be seen by expanding the \lstd\ in the pion fields. For example,
the kinetic term $\Tr\, \partial_\mu \Sigma\, \partial^\mu \Sigma^\dagger$ has
a coefficient of order
\eqn\XIIii{
f^2\lchi^2 \left(\bfrac{\partial}{\lchi}\right)^2 \sim f^2,
}
the four derivative term $\Tr\,
\partial_\mu \Sigma\, \partial^\mu \Sigma^\dagger \partial_\nu \Sigma\,
\partial^\nu \Sigma^\dagger$ has a coefficient of order
\eqn\XIIiii{
f^2\lchi^2 \left(\bfrac{\partial}{\lchi}\right)^4 \sim
\bfrac{f^2}{\lchi^2}\sim
\bfrac1{16\pi^2}, \ \ {\rm etc.},
}
which agrees with the earlier estimates.

\titlea{12}{Explicit Symmetry Breaking}
The light quark masses explicitly break the chiral $SU(3)_L\times SU(3)_R$
symmetry of the QCD Lagrangian. The quark mass term in the QCD Lagrangian is
\eqn\lmass{
\CL_m=-\bar \psi_L M \psi_R + h.c.,
} where
\eqn\mmatrix{
M = \left[\matrix{ m_u &0 & 0\cr 0 & m_d & 0 \cr 0 & 0 &m_s \cr} \right], } is
the quark mass matrix. The mass term $\CL_m$ can be treated as chirally
invariant if $M$ is an external field that transforms as
\eqn\mtrans{
M\rightarrow L M R^\dagger } under chiral $SU(3)_L\times SU(3)_R$. The
symmetry
breaking terms in the chiral Lagrangian are terms that are invariant when $M$
has the transformation rule \mtrans. The symmetry is then explicitly
broken
when $M$ is fixed to have the value \mmatrix. The lowest order term in the
effective Lagrangian to first order in $M$ is
\eqn\mterm{
\CL_m=\mu\bfrac {f^2} 2 \tr\left(\Sigma^\dagger M + M^\dagger \Sigma\right),
} which breaks the degeneracy of the vacuum and picks out a particular
orientation for $\Sigma$. All vacua $\Sigma={\rm constant}$ are no longer
degenerate, and $\Sigma=1$ is the lowest energy state. Expanding in small
fluctuations about $\Sigma=1$ gives
\eqn\XIIiv{
\CL_m = - 2 \mu \tr M \pi^2.
}
Substituting \mmatrix\ and \pifield\ for $M$ and $\pi$ and evaluating the
trace gives
\eqn\pimasses{\eqalign{
M^2_{\pi^\pm} &= \mu\left(m_u+m_d\right)+\Delta M^2,\cr M^2_{K^\pm} &=
\mu\left(m_u+m_s\right)+\Delta M^2,\cr M^2_{K^0,\bar K^0} &=
\mu\left(m_d+m_s\right), }} and the $\pi^0$, $\eta$ mass matrix
\eqn\etapimatrix{
\mu\left[\matrix{m_u+m_d&\bfrac{m_u-m_d}{\sqrt3}\cr
\bfrac{m_u-m_d}{\sqrt3}&\bfrac13\left(m_u+m_d+4m_s\right)\cr}\right].
} To first order in the isospin-breaking parameter $m_u-m_d$, the matrix
\etapimatrix\ has eigenvalues
\eqn\etapieigen{\eqalign{
M^2_{\pi^0} &= \mu\left(m_u+m_d\right),\cr M^2_{\eta} &=
\bfrac\mu3\left(m_u+m_d+4m_s\right).  }} There is an isospin breaking
electromagnetic contribution to the charged Goldstone boson masses $\Delta
M^2$
(included in \pimasses), which is comparable in size to the isospin
breaking from $m_u-m_d$. To lowest order in $SU(3)$ breaking, $\Delta M^2$ is
equal for $\pi^\pm$ and $K^\pm$, and vanishes for the neutral mesons. The
absolute values of the quark masses can not be determined from the meson
masses, because they always occur in the combination $\mu m$, and $\mu$ is an
unknown parameter. However, the meson masses can be used to obtain quark mass
ratios. From \pimasses--\etapimatrix\ one gets
\eqn\mumd{
\bfrac{m_u}{m_d} = {M^2_{K^+}-M^2_{K^0}+2M^2_{\pi^0}-M^2_{\pi^+}
\over M^2_{K^0}-M^2_{K^+}+M^2_{\pi^+}},
}
\eqn\msmd{
\bfrac{m_s}{m_d} = {M^2_{K^0}+M^2_{K^+}-M^2_{\pi^+}\over
M^2_{K^0}-M^2_{K^+}+M^2_{\pi^+}}, } and the Gell-Mann--Okubo formula
\eqn\gmo{
4 M^2_{K^0} = 3 M^2_\eta + M^2_\pi.  } Substituting the measured meson masses
gives the lowest order values
\eqn\mumdms{
\bfrac{m_u}{m_d} = 0.55,\qquad \bfrac{m_s}{m_d} =20.1.
} and $0.99\ {\rm GeV}^2=0.92\ {\rm GeV}^2$ for the Gell-Mann--Okubo formula.

There is an ambiguity in extracting the light quark masses at second order in
$M$. The matrices $M$ and $(\det M) M^{\dagger -1}$ both have the same
$SU(3)_L
\times SU(3)_R$ transformation properties, and are indistinguishable in the
chiral Lagrangian. One has an ambiguity of the form
\eqn\mambig{
M\rightarrow M + \lambda\left(\det M\right) M^{\dagger -1} } in the quark mass
matrix at second order in $M$. This transformation can be written explicitly
as
\eqn\mambigt{
\left[\matrix{ m_u &0 & 0\cr 0 & m_d & 0 \cr 0 & 0 &m_s \cr} \right]
\rightarrow
\left[\matrix{ m_u+\lambda m_d m_s &0 & 0\cr
0 & m_d+\lambda m_u m_s & 0 \cr 0 & 0 &m_s+\lambda m_um_d \cr} \right].  } One
cannot determine the light quark mass ratios to second order using chiral
perturbation theory alone, because of the ambiguity \mambig. This
ambiguity
can be numerically significant for the ratio $m_u/m_d$, since it produces an
effective $u$-quark mass of order $m_d m_s/\lchi\sim m_d m_K^2/\lchi^2
\sim 0.3 m_d$. The value of $m_u$ is very important, because $m_u=0$ solves
the
strong CP problem. The second order term $(\det M) M^{\dagger -1}$ produces an
effective $u$-quark mass that is indistinguishable from $m_u$ in the chiral
Lagrangian.  Various estimates of the ratio $m_u/m_d$ from different processes
(e.g. meson masses, baryon masses $\eta\rightarrow 3\pi$) all tend to give the
ratio $m_u/m_d\sim 0.56$, so $m_u$ can only be zero if second-order effects in
$M$ were of the same size in different processes. One way this could occur is
if instantons effects at the scale $\mu\sim1~{\rm GeV}$ were important. An
instanton produces an effective operator of the form $(\det M) M^{\dagger
-1}$. If instantons at $\mu\sim1~{\rm GeV}$ are important, they would lead to
an effective mass matrix in the chiral Lagrangian of the form $M_{\rm eff} =
M+
\lambda\det M M^{-1}$, which would be the same for all processes. This
produces
$(m_u/m_d)_{\rm eff}\sim 0.56$ in all processes, while still having
$m_u=0$. The only way to distinguish $m_u$ from $(m_u)_{\rm eff}$ is to do a
reliable computation that relates the QCD Lagrangian directly to the chiral
Lagrangian.

An on-shell particle has $p^2=M^2$. Since the meson mass-squared is
proportional to the quark mass $M$, the quark matrix $M$ counts as two powers
of $p$ for chiral power counting, i.e. terms in $\CL_2$ contain two powers of
$p$ or one power of $M$, terms in $\CL_4$ contain four powers of $p$, two
powers of $p$ and one power of $M$, or two powers of $M$, etc. One can then
show that the power counting arguments derived earlier still hold for the
effective Lagrangian, including symmetry breaking.

\titlea{13}{$\pi$-$\pi$ Scattering}
We now have all the pieces necessary to compute the $\pi$-$\pi$ scattering
amplitude near threshold. The full chiral Lagrangian to order $p^2$ is
\eqn\ltwom{
\CL_2 = \bfrac{f^2}{4} \tr \partial_\mu\Sigma\partial^\mu\Sigma^\dagger +
\mu\bfrac {f^2}2 \tr \left(\Sigma^\dagger M + M^\dagger \Sigma\right).
} Expanding this to fourth order in the pion fields gives
\eqn\ltwomexp{
\CL_2=\bfrac1{3f^2}\Tr \left[\pi,\partial_\mu\pi\right]^2+
\bfrac23 \mu \Tr M \pi^4.
} The $\pi$-$\pi$ scattering amplitude has two contributions, one from the
kinetic term and the other from the mass term. Adding the two contributions
reproduces the result of Weinberg. The details are left as a homework problem.

The $\pi$-$\pi$ scattering amplitude at order $p^4$ has two contributions, the
one loop diagram Fig.~\fig\pipiloop{}(a) involving only the lowest order
Lagrangian,
and a tree graph Fig.~\pipiloop(b) with terms from $\CL_4$. The answer has the
form
\eqn\genform{
{A\over 16\pi^2}\ p^4\, \log p^2/\mu^2 + L(\mu)\ p^4, } where the first term
is
from the loop diagram, and the second term is the tree graph contribution from
$\CL_4$. The coefficient $A$ of the loop graph is completely determined, since
there are no unknown parameters in $\CL_2$. The loop graph must have a
logarithmic term, the so-called chiral logarithm. When $s>4m^2_{\pi}$, the
$\pi$-$\pi$ scattering amplitude must have an imaginary part from the physical
$\pi\pi$ intermediate state, by unitarity. The imaginary part is generated by
the chiral logarithm. When $s>4m^2_{\pi}$, the argument of the logarithm
changes sign, and one gets an imaginary part since $\log (-\abs r) = \log \abs
r + i\pi$. The imaginary part is completely determined by the tree level graph
of order $p^2$, so that the chiral logarithm has a known coefficient. The tree
level terms in $\CL_4$ are known as low energy constants or
counterterms.\fonote{There are two parts to $\CL_4$. There is the infinite
part of the coefficient (of order $1/\epsilon$ in dimensional regularization),
which is used to cancel divergences from loops using
vertices in $\CL_2$, and the finite part which affects measurable quantities.
The infinite parts are usually never discussed explicitly, and the finite parts
are called counterterms.} The total scattering
amplitude is $\mu$ independent, so the counterterms satisfy the
renormalization
group equation
\eqn\rgct{
\mu {d\over d\mu} L(\mu) = {A\over 8\pi^2}.
} The naive dimensional analysis argument discussed earlier is the statement
that the counterterm $L(\mu)$ is typically at least as big as the anomalous
dimension $A/8\pi^2$.

\begfig 0 cm
\centerline{\hbox{
\epsfxsize=6cm\epsffile{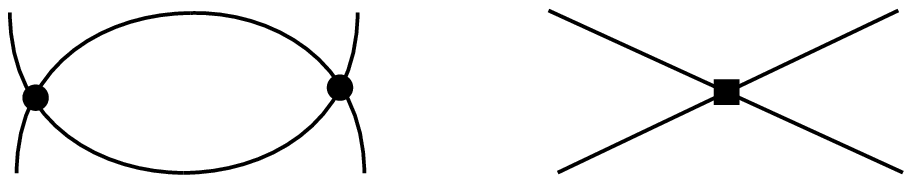}}}
\figure{\pipiloop}{Diagrams contributing to $\pi-\pi$ scattering to
order $p^4$. The solid dot represents interaction vertices from
$\CL_2$, and the solid square represents interactions from
$\CL_4$}
\endfig

A generic chiral perturbation theory amplitude has the form \genform.
There
is a chiral logarithm and some counterterms. If one works in a systematic
expansion in powers of $p$, the chiral logarithm is determined completely in
terms of lower order terms in the Lagrangian. The counterterms involve
additional unknown parameters. There are three main approaches used in the
literature to extract useful information from \genform:
\medskip
\item{1.} One can hope that the chiral logarithm is numerically more
important
than the counterterm, when one picks a reasonable renormalization point such
as
$\mu\sim1$~GeV. This is formally correct, since $p^4\log p^2/\mu^2 \gg$ $p^4$
in the limit $p\rightarrow 0$. However, in practical examples, $p^2$ is of
order $m_\pi^2$ or $m_K^2$, and the logarithm is $-3.9$ and $-1.4$, which is
not very large (especially for the $K$). Nevertheless, the chiral logarithm
provides useful information. For example, the correction to $f_K/f_\pi$ has
the
form
\eqn\fkfpi{
\bfrac{f_K}{f_\pi} = 1 - \bfrac{3M_K^2}{64\pi^2 f^2}\, \log M_K^2/\mu^2 +
L(\mu).  } Setting $\mu\sim\lchi$, and neglecting the counterterm gives
$f_K/f_\pi=1.19$, compared with the experimental value of $1.2$. The chiral
logarithm contribution alone gives a reasonable estimate of the size of the
correction (this is just naive dimensional analysis at work), but it also gets
the sign correct. The chiral logarithms are also useful in comparing numerical
QCD calculations in the quenched approximation, which do not have the full
chiral logarithms, with experimental data.
\item{2.} The systematic approach which has been used by Gasser and
Leutwyler
is to write down the most general Lagrangian to order $p^4$, which contains
eight counterterms. This is used to compute $N>8$ different processes, so that
all the counterterms are determined, and one has non-trivial predictions for
the remaining $N-8$ amplitudes. This procedure has been more-or-less completed
in the meson sector to order $p^4$, and the results are in good agreement with
experiment. At order $p^6$, there are over 100 terms in the Lagrangian.
\item{3.} The third method is to find a process for which there is no
counterterm. Typically, this occurs for electromagnetic processes involving
neutral particles, such as $K^0_S\rightarrow \gamma\gamma$. Since there is no
counterterm, the loop graph must be finite, but it can be non-zero. For
example, the leading contribution to $K^0_S\rightarrow \gamma\gamma$ is from
the loop graph Fig.~\fig\figkshort{kshort}, and gives an amplitude of order
$p^2$. There are no counterterms for this process at this order. The amplitude
at order $p^2$ is in good agreement with the experimental branching ratio for
this process.

\begfig 0 cm
\centerline{\hbox{
\epsfxsize=4cm\epsffile{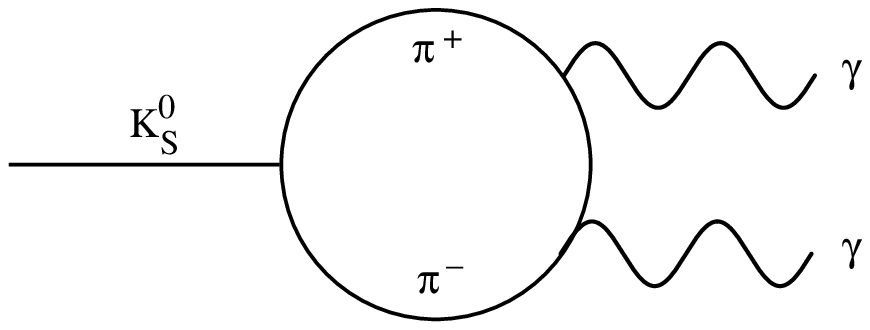}}}
\figure{\figkshort}{Leading contribution to $K^0_S \rightarrow
\gamma \gamma$}
\endfig

\titlea{14}{Chiral Perturbation Theory for Matter Fields}
Chiral perturbation theory can also be applied to the interactions of the
Goldstone bosons with all other particles, which are generically referred to as
matter fields. The matter fields (baryon, heavy mesons, etc.) transform as
irreducible representations of $SU(3)_V$, but do not form representations of
chiral $SU(3)_L\times SU(3)_R$. To discuss the interactions of matter fields,
it is more convenient to use the $\xi$-basis of Sect.~11. We will
consider
the interactions of the pions with the spin-1/2 baryon octet. The
generalization to other matter fields will be obvious. The CCWZ prescription
for matter fields such as the baryon is that under a $SU(3)_L\times SU(3)_R$
transformation, the transformation law is
\eqn\btrans{
B\rightarrow U B U^\dagger, } where $U$ is implicitly defined in terms of $L$
and $R$ in \xitrans, and the octet of baryon fields is
\eqn\bmatrix{
\left[\matrix{\bfrac1{\sqrt2}\Sigma^0 +
\bfrac1{\sqrt6}\Lambda&\Sigma^+&p\cr\Sigma^-&-\bfrac1{\sqrt2}\Sigma^0 +
\bfrac1{\sqrt6}\Lambda&n\cr\Xi^-&\Xi^0&-\bfrac2{\sqrt6}\Lambda}\right].
} Under a $SU(3)_V$ transformation, $L=R=U$, and the baryon transforms as an
$SU(3)_V$ adjoint. Any transformation that reduces to the adjoint
transformation law for $SU(3)_V$ transformations is acceptable. For example,
one can choose
\eqn\XIIv{
B \rightarrow L B L^\dagger,\qquad B \rightarrow U B R^\dagger,\qquad etc.
}
The different choices are all equivalent, and correspond to redefining the
baryon field. For example, if $B$ has the transformation law \btrans, then
$\xi B\xi^\dagger$ and $B\xi$ transform as $B \rightarrow L B L^\dagger$ and
$B
\rightarrow U B R^\dagger$ respectively.

The baryon chiral Lagrangian is the most general invariant Lagrangian written
in terms of $B$ and $\xi$. In writing the Lagrangian, it is convenient to
introduce the definitions
\eqn\avdef{\eqalign{
A^\mu &= \frac i 2 \left(\xi\partial^\mu \xi^\dagger - \xi^\dagger
\partial^\mu \xi\right) = \bfrac{\partial^\mu \pi}f + \ldots,\cr
V^\mu &= \frac 1 2 \left(\xi\partial^\mu \xi^\dagger + \xi^\dagger
\partial^\mu \xi\right)=\bfrac{1}{2f^2}\left[\pi,\partial^\mu \pi\right]
+\ldots, }} which transform as
\eqn\atrans{
A^\mu \rightarrow U A^\mu U^\dagger, } and
\eqn\vtrans{
V^\mu\rightarrow U V^\mu U^\dagger - \partial^\mu U U^\dagger, } under
$SU(3)_L\times SU(3)_R$. The covariant derivative on baryons is defined by
\eqn\covder{
D^\mu B = \partial^\mu B + \left[V^\mu,B\right], } which transforms as
\eqn\covtrans{
D^\mu B \rightarrow U\, D^\mu B\, U^\dagger.  } The most general baryon
Lagrangian to order $p$ is
\eqn\lbi{
\CL = \tr \bar B \left(i \dsl-m_B\right) B + D \tr \bar B \gamma^\mu
\gamma_5 \left\{A_\mu,B\right\} + F \tr \bar B \gamma^\mu
\gamma_5 \left[A_\mu,B\right] + \CL_{\xi},
} where $\CL_{\xi}$ is the purely meson Lagrangian with $\Sigma = \xi^2$,
$m_B$
is the baryon mass, $\dsl$ is the covariant derivative \covder\
and $F$ and $D$ are the usual axial vector coupling
constants, with $g_A=F+D$.

The presence of the dimensionful parameter $m_B$ in the Lagrangian ruins the
power counting arguments necessary for a sensible effective field theory. Loop
graphs in baryon chiral perturbation theory will produce corrections of order
$m_B/\lchi\sim1$, so the entire chiral expansion breaks down. There is an
alternative formulation of baryon chiral perturbation theory that avoids this
problem. The idea is to expand the Lagrangian about nearly on-shell baryons,
so
that one has a Lagrangian that can be expanded in powers of $1/m_B$, and has
no
term of order $m_B$. The method used is similar to that used for heavy quark
fields in HQET. Instead of using the Dirac baryon field $B$, one uses a
velocity-dependent baryon field $B_v$, which is related to the original baryon
field $B$ by
\eqn\bvdef{
B_v(x) = \bfrac{1+\slash v}{2}\, B(x)\ e^{im_B v \cdot x}, } where $v$ is the
velocity of the baryon. In the baryon rest frame, $v=(1,0,0,0)$, and
\eqn\brest{
B_v(x) = \bfrac{1+\gamma^0}{2}\, B(x)\ e^{im_B t}, } which corresponds to
keeping only the particle part of the spinor, and subtracting the baryon mass
$m_B$ from all energies. In terms of the field $B_v$, the chiral Lagrangian is
\eqn\lvii{
\CL_v = \tr \bar B \left(iv\cdot D\right) B + D \tr \bar B \gamma^\mu
\gamma_5 \left\{A_\mu,B\right\} + F \tr \bar B \gamma^\mu
\gamma_5 \left[A_\mu,B\right] +\CO\left(\bfrac{1}{m_B}\right)+ \CL_{\xi}.
} The baryon mass term is no longer present, and the baryon Lagrangian now has
an expansion in powers of $1/m_B$. Note that the baryon chiral Lagrangian
starts at order $p$, whereas the meson Lagrangian starts at order $p^2$.

A similar procedure can be applied to other matter fields, not just to
baryons,
provided one can factor the common mass (such as $m_B$) out of the Feynman
graphs. For baryon chiral perturbation theory, this is possible because baryon
number is conserved, so one can remove a common mass $m_B$ from all baryons. A
similar method also works for hadrons containing a heavy quark, such as the
$B$
and $B^*$ mesons, because $b$-quark number is conserved by the strong
interactions, and the $B$ and $B^*$ are degenerate in the heavy quark limit.
It cannot be used for processes such as $\rho\rightarrow\pi \pi$, because the
$\rho$ mass $m_\rho$ turns into the pion energy in the final state.

The velocity-dependent Lagrangian $\CL_v$ has no dimensionful coefficients in
the numerator. This implies that the power counting arguments of an effective
field theory are valid. One has two expansion parameters, $1/m_B$ and
$1/\lchi$. The power counting rule \weini\ is now
\eqn\weinii{
D = 1 + 2L + \sum_k m_k\left(k-2\right) + \sum_k n_k\left(k-1\right), } where
$m_k$ is the number of vertices from the $p^k$ terms in the meson Lagrangian,
and $n_k$ is the number of vertices from the $p^k$ terms in the baryon
Lagrangian. The proof of this result is similar to \weini, and will be
omitted. The difference between the meson and baryon terms arises because the
meson propagator is $1/k^2$, whereas the baryon propagator is $1/k\cdot v$.

The naive dimensional analysis estimate \ndest\ is now
\eqn\ndab{
f^2\lchi^2\left(\bfrac{\pi}f\right)^n
\left(\bfrac{\partial}{\lchi}\right)^m
\left(\bfrac{B}{f\sqrt\lchi}\right)^r
} For example, the kinetic term $\bar B\left(i v\cdot D \right)B$ has a
coefficient of order
\eqn\XIIxvi{
f^2\lchi^2
\left(\bfrac{\partial}{\lchi}\right)^1\left(\bfrac{B}{f\sqrt\lchi}\right)^2
\sim 1,
}
and the four-baryon term $\bar B B\ \bar B B$ has a coefficient of order
\eqn\fourbary{
f^2\lchi^2\left(\bfrac{B}{f\sqrt\lchi}
\right)^4\sim\bfrac 1{f^2}.
}

Similar power counting arguments hold for all strongly interacting gauge
theories. For example, in tests for quark and lepton substructure, one uses
the
operator
\eqn\XIIvi{
\bfrac{4\pi}{\Lambda_{\rm ELP}^2}\ \bar q q\ \bar q q,
}
and places limits on $\Lambda_{\rm ELP}$. A quark field has the same power
counting
rules as a baryon field in baryon chiral perturbation theory. Comparing with
\fourbary, we see that
\eqn\XIIvii{
\Lambda_{\rm ELP} = \Lambda/\sqrt{4\pi},
}
where $\Lambda$ is the scale of the composite interactions defined by analogy
with the chiral scale $\lchi$: i.e. scattering amplitudes vary on a momentum
scale $\Lambda$.

\titleb{}{$\pi N$ Scattering}
A simple application of the baryon chiral Lagrangian is the computation of
$\pi
-N$ scattering amplitude at threshold, to order $p$. From \weinii, the
only
graphs which contribute are tree graphs which involve terms from the meson
Lagrangian at order $p^2$, and the baryon Lagrangian at order $p$. The two
diagrams which contribute are shown in Fig.~\fig\figpiN{pi N}. The pion-nucleon
vertex in Fig.~\figpiN(a) vanishes at threshold, since it is proportional to
$\bfm
p$. The two-$\pi$ nucleon vertex in Fig.~\figpiN(b) is
\eqn\piNvertex{
\bfrac {i}{2f^2} \bar B\left[\pi,\partial^\mu\pi\right] v^\mu B.
} The amplitude can be rewritten using $\pi=\pi^a T^a_B$, where $T^a_B$ are
the
flavor matrices in the baryon representation. The pions are in the adjoint
representation of flavor, so the flavor matrices acting on pions can be
written
in terms of the structure constants
\eqn\XIIxviii{
\left(T^c_\pi\right)_{ba} = i f_{abc}.
}
Using the commutation relations $\left[T^a_B,T^b_B\right]=if_{abc} T^c_B$, and
evaluating Fig.~\figpiN\ using the interaction \piNvertex\ gives the amplitude
\eqn\piNamp{
\CA =- \bfrac i {f^2}\ M_\pi\ T_\pi \cdot T_B,
} where we have used $E=M_\pi$ for the energy of the pion.  Changing from the
non-relativistic normalization of baryon states to the relativistic
normalization (where the states are normalized to $2E$) gives the
Weinberg-Tomozawa formula for the pion-nucleon scattering amplitude
\eqn\weintom{
\CA =- \bfrac {2i} {f^2}\ M_B M_\pi\ \left(T_\pi \cdot T_B\right).
} $T_\pi\cdot T_B=1/2\left[(T_\pi+T_B)^2-T_\pi^2-T_B^2\right]=1/2\left[
I(I+1)-2-3/4\right]$, so that $T_\pi\cdot T_B$ is $-1$ in the isospin-1/2
channel and $1/2$ in the isospin-3/2 channel.

\begfig 0 cm
\centerline{\hbox{
\epsfxsize=12cm\epsffile{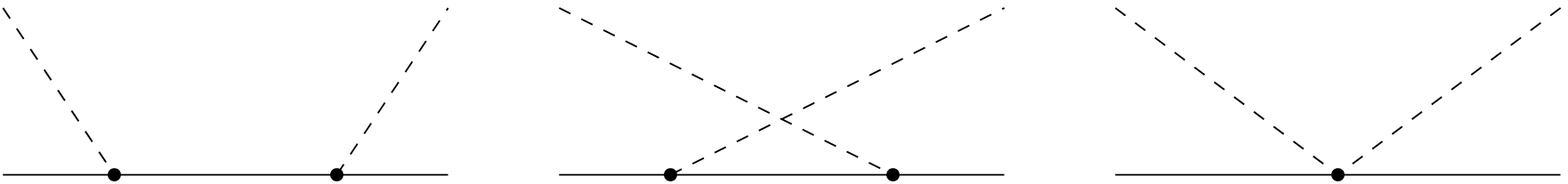}}}
\figure{\figpiN}{Contributions to $\pi N$ scattering at order
$p$}
\endfig

\titleb{}{Non-Analytic Terms}
The chiral Lagrangian for matter fields can be used to compute loop
corrections. The matter field Lagrangian has an expansion in powers of $p$,
whereas the Goldstone boson Lagrangian had an expansion in powers of
$p^2$. Consequently, loop integrals for matter field interactions can have
either even or odd dimension. The even dimensional integrals have the same
structure as for the mesons, and lead to non-analytic terms of the form
$(M^{2r}/16\pi^2 f^2) \log M^2/\mu^2$, where $M$ is a $\pi$, $K$ or $\eta$
mass, and $r$ is an integer. The $\mu$ dependence is cancelled by a
corresponding $\mu$ dependence in
a higher order term in the Lagrangian. The odd dimensional integrals lead to
non-analytic terms of the form $(M^{2r+1}/16\pi f^2)$, where $r$ is an
integer.
These odd-dimensional terms do not have a multiplying logarithm, since the
$\mu$ dependence cannot be absorbed by a higher dimension operator in the
effective Lagrangian. The operator would have to have the form
$M^{2r+1}$ which is proportional to $m_q^{r+1/2}$, where $m_q$ is the quark
mass. Such an operator cannot exist in the Lagrangian, since it contains a
square-root of the quark mass matrix. Also note that the odd-dimensional
integrals have one less power of $\pi$ in the denominator.

\titlea{15}{Chiral Perturbation Theory for Hadrons Containing a Heavy Quark}
Chiral perturbation theory can also be applied to hadrons containing a heavy
quark. The hadrons are treated as matter fields, and one writes down the most
general possible Lagrangian consistent with the chiral symmetries, as for the
spin-1/2 baryons. In additional, one can constrain some of the terms in the
Lagrangian using heavy quark symmetry. As a simple example, consider the
interaction of the pseudoscalar and vector mesons $B$ and $B^*$ (or $D$ and
$D^*$) with pions. These mesons can be treated using the velocity dependent
formulation, since $b$-quark number is conserved by the strong interactions,
and the $B$ and $B^*$ are degenerate in the heavy quark limit. It is
conventional to combine $B$ and $B^*$ into a single field $H$ defined by
\eqn\XVi{
H = \bfrac{1+\slash v}{2}\left[B^*_\mu \gamma^\mu - B\gamma_5\right],
}
where $B$ and $B^*$ are column vectors which contain the states $b\bar u$,
$b\bar d$, and $b\bar s$. The transformation law for $H$ under heavy quark
spin
symmetry transformation $S_Q$ and $SU(3)_V$ flavor symmetry transformation $U$
is
\eqn\XVii{
H \rightarrow S_Q H U^\dagger.
}
The most general Lagrangian consistent with these symmetries to order $p$ is
\eqn\hqetl{
\CL = \tr \bar H \left(iv\cdot D\right) H + g \tr \bar H H \gamma^\mu\gamma_5
A_\mu, } where
\eqn\hcovder{
D^\mu H = \partial^\mu H - H V^\mu.  } There is only a single coupling
constant
$g$ which appears to this order, so the $B B^*\pi$ and $B^* B^*\pi$ couplings
are related to each other by the heavy quark spin symmetry. The other possible
interaction term,
\eqn\XViii{
 \tr \bar H \gamma^\mu\gamma_5 H A_\mu
}
is forbidden by heavy quark spin symmetry, and is suppressed by one power of
$1/m_Q$. This term splits the $B B^*\pi$ and $B^* B^*\pi$ couplings at order
$1/m_Q$.

The chiral Lagrangian \hqetl\ can be used to compute corrections to
various
quantities for heavy hadrons. For example, one can show that
\eqn\fratio{
\bfrac{f_{B_s}}{f_B} = 1 - \frac56\left(1+3g^2\right){M_K^2\over
16 \pi^2 f^2}\log {M_K^2\over\mu^2}, } and
\eqn\bratio{
\bfrac{B_{B_s}}{B_B} = 1 - \frac23\left(1-3g^2\right){M_K^2\over
16 \pi^2 f^2}\log {M_K^2\over\mu^2}, } where $f_B$ and $B_B$ are the decay
constant for $B$ decay and the bag constant for $B^0-\bar{B}^0$ mixing
respectively. Further applications can be found in the literature.

\titlea{16}{Acknowledgments}
I would like to thank Gabriela~Barenboim, C.~Glenn~Boyd, Richard~F.~Lebed,
Ira~Z.~Rothstein, and Oscar~Vives
for helpful comments
on the manuscript.  This work was supported in part by a Department of Energy
grant DOE-FG03-90ER40546, by a Presidential Young Investigator award from
the National Science Foundation, PHY-8958081, and by a ``Profesor
Visitante IBERDROLA de Ciencia y Tecnolog\'{\i}a'' position at
the Departmento de F\'{\i}sica Te\'orica of the University of Val\`encia.

\begref{References}{}
Here are a few references which might be useful for students
interested in learning more about the subject.
\bigskip

\global\newcount\itemno \global\itemno=0
\def\nitem#1{\global\advance\itemno by1\item{\the\itemno.}}

\nitem{1.} The renormalization group and critical phenomena:
\itemitem\ref Ma,\ts S.-K. (1982): {\it Modern Theory of Critical Phenomena}
(Benjamin)
\itemitem\ref Wilson,\ts K., Kogut,\ts J. (1974): Phys.\ts Rep.
{\bf 12}, 75
\smallskip

\nitem{2.} Some references on effective field theories:
\itemitem\ref Georgi,\ts H. (1984): {\it Weak Interactions and Modern Particle
Theory} (Benjamin)
\itemitem\ref Hall,\ts L.J. (1981): Nucl.\ts Phys. {\bf B178}, 75
\itemitem\ref Kaplan,\ts D.B. (1995): e-print {\tt nucl-th/9506035}
\itemitem\ref Polchinski,\ts J. (1992): e-print {\tt hep-th/9210046}
\itemitem\ref Weinberg,\ts S. (1980): Phys.\ts Lett. {\bf 91B}, 51
\itemitem\ref Witten,\ts E. (1977): Nucl.\ts Phys. {\bf B122}, 109
\smallskip

\nitem{3.} The sine-Gordon--Thirring duality:
\itemitem\ref Coleman,\ts S. (1975): Phys.\ts Rev. {\bf D11}, 2088
\smallskip

\nitem{3.} The Appelquist-Carazzone theorem:
\itemitem\ref Appelquist,\ts T., Carazzone,\ts J. (1975):
Phys.\ts Rev. {\bf D11}, 2856
\smallskip

\nitem{5.} Weak interactions in the six-quark model:
\itemitem\ref Gilman,\ts F.J., Wise,\ts M.B. (1979): Phys.\ts Rev.
{\bf D20}, 2392
\itemitem\ref Gilman,\ts F.J., Wise,\ts M.B. (1983): Phys.\ts Rev.,
{\bf D27}, 1128
\itemitem\ref Shifman,\ts M., Vainshtein,\ts A., Zakharov,\ts V. (1977):
Nucl.\ts Phys. {\bf B120}, 316
\itemitem\ref Vainshtein,\ts A., Zakharov,\ts V., Shifman,\ts M. (1975):
JETP\ts Lett. {\bf 22}, 55
\smallskip

\nitem{6.} Non-local effective actions:
\itemitem\ref Bhansali,\ts V., Georgi,\ts H. (1992): e-print
{\tt hep-ph/9205242}
\smallskip

\nitem{7.} Simplifying the effective Lagrangian using equations of motion:
\itemitem\ref Georgi,\ts H. (1991): Nucl.\ts Phys. {\bf B361}, 339
\itemitem\ref Politzer,\ts H.D. (1980): Nucl.\ts Phys. {\bf B172}, 349
\smallskip

\nitem{8.} The CCWZ formalism:
\itemitem\ref Callan,\ts C., Coleman,\ts S., Wess,\ts J., Zumino,\ts B.
(1969): Phys.\ts Rev. {\bf 177}, 2247
\itemitem\ref Coleman,\ts S., Wess,\ts J., Zumino,\ts B. (1969):
Phys.\ts Rev. {\bf 177}, 2239
\smallskip

\nitem{9.} A summary of many of the ideas on chiral perturbation theory:
\itemitem\ref Weinberg,\ts S. (1979): Physica {\bf A96}, 327
\smallskip

\nitem{10.} Naive dimensional analysis:
\itemitem\ref Manohar,\ts A., Georgi,\ts H. (1984):
Nucl.\ts Phys {\bf B234}, 189
\smallskip

\nitem{11.} The existence of non-analytic terms in the chiral expansion:
\itemitem\ref Li,\ts L.-F., Pagels,\ts H. (1972): Phys.\ts Rev. {\bf D5}, 1509
\smallskip

\nitem{12.} The meson Lagrangian to order $p^4$:
\itemitem\ref Gasser,\ts J., Leutwyler,\ts H. (1984):
Ann.\ts Phys. {\bf 158}, 142
\itemitem\ref Gasser,\ts J., Leutwyler,\ts H. (1985): Nucl.\ts Phys.
{\bf B250}, 465
\smallskip

\nitem{13.} Quark masses:
\itemitem\ref Gasser,\ts J., Leutwyler,\ts H. (1982):
Phys.\ts Rept. {\bf 87}, 77
\itemitem\ref Kaplan,\ts D.B., Manohar,\ts A.V. (1986):
Phys.\ts Rev.\ts Lett. {\bf 56}, 2004
\itemitem\ref Weinberg,\ts S. (1977): Trans.\ts N.Y.\ts Acad.\ts Sci.
{\bf 38}, 185
\smallskip

\vbox{
\nitem{14.} Baryon chiral perturbation theory:
\itemitem\ref Bijnens,\ts J., Sonoda,\ts H., Wise,\ts M.B. (1985): Nucl.\ts
Phys. {\bf B261}, 185
\itemitem\ref Gasser,\ts J., Sainio,\ts M.E., Svarc,\ts A. (1988): Nucl.\ts
Phys. {\bf B307}, 779
\itemitem\ref Jenkins,\ts E., Manohar,\ts A.V. (1991a): Phys.\ts Lett.
{\bf B255}, 558
\itemitem\ref Jenkins,\ts E., Manohar,\ts A.V. (1991b): Phys.\ts Lett.
{\bf B259}, 353
\itemitem\ref Langacker,\ts P., Pagels,\ts H. (1974): Phys.\ts Rev.
{\bf D10}, 2904
\smallskip
}
\nitem{15.} A good reference to recent developments in chiral perturbation
theory:
\itemitem\ref Mei\ss ner,\ts U., Ed. (1992): {\it Effective Field Theories
of the Standard Model} (World Scientific, Singapore)
\smallskip

\nitem{16.} Chiral perturbation theory for hadrons containing a heavy quark:

\itemitem\ref Burdman,\ts G., Donoghue,\ts J. (1992): Phys.\ts Lett. {\bf
B280},
287
\itemitem\ref Wise,\ts M.B. (1992): Phys.\ts Rev. {\bf D45}, 2188
\itemitem\ref Yan,\ts T.M. et al. (1992): Phys.\ts Rev. {\bf D46}, 1148
\smallskip

\nitem{17.} Some applications of heavy hadron chiral perturbation theory:
\itemitem\ref Grinstein,\ts B. et al. (1992): Nucl.\ts Phys. {\bf B380}, 376
\endref
\bye